\documentclass[english,aps,preprint]{revtex4}
\usepackage[T1]{fontenc}
\usepackage[latin9]{inputenc}
\usepackage{array}
\usepackage{prettyref}
\usepackage{float}
\usepackage{units}
\usepackage{amsbsy}
\usepackage{graphicx}
\usepackage{amssymb}
\usepackage{esint}

\makeatletter

\providecommand{\tabularnewline}{\\}

\@ifundefined{textcolor}{}
{%
 \definecolor{BLACK}{gray}{0}
 \definecolor{WHITE}{gray}{1}
 \definecolor{RED}{rgb}{1,0,0}
 \definecolor{GREEN}{rgb}{0,1,0}
 \definecolor{BLUE}{rgb}{0,0,1}
 \definecolor{CYAN}{cmyk}{1,0,0,0}
 \definecolor{MAGENTA}{cmyk}{0,1,0,0}
 \definecolor{YELLOW}{cmyk}{0,0,1,0}
 }

\makeatother

\usepackage{babel}

\begin{document}

\title{The Cosmological Constant as a Ghost of Inflaton}

\author{Yu-Chung Chen}

\email{yuchung.chen@gmail.com}

\address{Department of Physics, National Taiwan University, Taipei 10617,
Taiwan, R.O.C.}
\begin{abstract}
The cosmological constant (term) is the simplest way, presently known,
to illustrate the accelerating expansion of the universe. However,
because of/despite its simple appearance, there is much confusion
surrounding its essence. Theorists have been asking questions for
years: Is there a mechanism to explain this term? Is it really a constant
or a variable? Moreover, it seems that we have created a huge gulf
separating the theories of inflation and accelerating expansion. Can
we eliminate such an uncomfortable discontinuity?

In this paper, we will journey to see the growth of the universe from
the very beginning of inflation. To simplify our discussion, we will
briefly \textquotedblleft{}turn off\textquotedblright{} the effects
of real and dark matter and shall use inflaton (a classical scalar
field) dynamics with a time-varying inflaton potential $V\left(\phi,t\right)$
as the screen to watch this process. Relying on these conditions,
we propose a non-traditional method of obtaining the solution of scale
factor $\mathrm{R}\left(t\right)$, which is only dependent on $\dot{\phi}^{2}$,
and discover that the term $\nicefrac{\ddot{\mathrm{R}}\left(t\right)}{\mathrm{R}\left(t\right)}$
will be a constant after kinetic inflaton $\dot{\phi}$ is at rest.
This result can be regarded as the effective cosmological constant
phenomenally. Moreover, we will also \textquotedblleft{}rebuild\textquotedblright{}
$V\left(\phi,t\right)$, realize its evolutionary process and then,
according to the relationship between $V\left(\phi,t\right)$ and
$\dot{\phi}^{2}$, it will be possible to smoothly describe the whole
evolution of the universe from the epoch of inflation. Therefore,
the implications of our findings will mean that the gulf between theories
will disappear. Lastly, we will also see how the formula could provide
a framework for solving the old and new cosmological constant problems
as well as much more besides.
\end{abstract}
\maketitle

\section{introduction}

According to observational data \cite{key-1,key-2,key-3,key-4,key-5,key-6},
the major components required to build the universe as we see it today
are 73\% dark energy, 22\% dark matter, 5\% observable matter and
about 0.008\% radiation. \textquotedbl{}Dark energy\textquotedbl{}
is introduced theoretically to explain the accelerating expansion
of the universe and, as the word \textquotedblleft{}dark\textquotedblright{}
implies, only few of its properties are known. Firstly, for example,
according to the 1st Friedmann equation without the cosmological constant
\begin{equation}
\frac{\ddot{\mathrm{R}}}{\mathrm{R}}=-\frac{4\pi G}{3c^{2}}\left(\varepsilon+3p\right),\label{eq:1}\end{equation}
the equation of state $\omega\equiv\nicefrac{p}{\varepsilon}<\nicefrac{-1}{3}$
(where $\varepsilon$ is the energy density and $p$ is the pressure)
should be satisfied in order to make the \textquotedbl{}anti-gravity\textquotedbl{}
$\ddot{\mathrm{R}}>0$ become possible on the large scales of the
universe; secondly, the repulsive properties of dark energy require
its distribution to be highly homogenous and isotropic; thirdly, observations
show its density to be roughly $\unitfrac[10^{4}]{eV}{cm^{3}}$ \cite{key-7};
and finally there is still no evidence to suggest that it interacts
with matter through any of the fundamental forces other than gravity.
Up to the present moment, many dark energy models \cite{key-35,key-36,key-37,key-38,key-39,key-40,key-41}
have been proposed and we should not, of course, forget the simplest
one which was introduced by Einstein in 1917 \cite{key-8}: \begin{equation}
R_{\mu\nu}-\lambda g_{\mu\nu}=-\frac{8\pi G}{c^{4}}\left(T_{\mu\nu}-\frac{1}{2}g_{\mu\nu}T\right).\label{eq:2}\end{equation}
Here, the term $\lambda$ is the one that Einstein famously described
as \textquotedbl{}the biggest blunder of (his) life.\textquotedbl{}
Marvelously, however, the cosmological constant has gone on to become
a charming topic of cosmology and fundamental physics today \cite{key-10,key-42,key-43,key-44,key-45,key-46}.
Indeed, the mere history of the topic informs us of how strange of
the cosmological constant is. Based on his belief in Mach's principle,
in 1917 Einstein inserted the cosmological term $\lambda$ into \prettyref{eq:2}
so as to keep the universe static. Soon, de Sitter \cite{key-9} proposed
another static solution controlled only by $\lambda$, \begin{equation}
ds^{2}=\frac{1}{\cosh^{2}\left(\mathcal{H}r\right)}\left\{ c^{2}dt^{2}-dr^{2}-H^{-2}\sinh^{2}\left(\mathcal{H}r\right)\left[d\theta^{2}+\sin^{2}\theta d\varphi^{2}\right]\right\} ,\label{eq:3}\end{equation}
where the corresponding conditions are $\varepsilon=p=0$ and $\lambda=\frac{3}{\mathrm{R}^{2}}=3\mathcal{H}^{2}$
($\mathrm{R}$ is the radius of a 3-sphere universe). Several years
later, Weyl pointed out that a test body on de Sitter\textquoteright{}s
metric would display a redshift because the term $\Gamma_{tt}^{r}=-c^{2}\mathcal{H}\tanh\left(\mathcal{H}r\right)\neq0$
would give a redshift $z\simeq\mathcal{H}r\ll1$ \cite{key-10}. Therefore,
even though Hubble's discovery \cite{key-11} was not yet published,
Einstein mailed Wely in 1923 to give his reaction: \textquotedblleft{}\emph{If
there is no quasi-static world, then away with the cosmological term!}\textquotedblright{}
\cite{key-10}

However, the cosmological term can not be abandoned so easily. According
to quantum field theory, anything that contributes to the energy density
of vacuum must act exactly like a cosmological constant. To repeat
Weinberg's elegant report \cite{key-10}, the vacuum energy-momentum
tensor must take the form\begin{equation}
\left\langle T_{\mu\nu}\right\rangle _{\mathrm{vac}}=\left\langle \varepsilon_{\mathrm{vac}}\right\rangle g_{\mu\nu}\label{eq:4}\end{equation}
to obey the Lorentz invariance (where we set $g_{00}=c^{2}$) and
the 2nd Fridemann equation in flat spacetime\begin{equation}
H^{2}=\left(\frac{\mathrm{\dot{R}}}{\mathrm{R}}\right)^{2}=\frac{8\pi G}{3c^{2}}\left\langle \varepsilon_{\mathrm{tot}}\right\rangle +\frac{\lambda c^{2}}{3},\label{eq:5}\end{equation}
where the total density can be separated into the ordinary part and
the vacuum part as

\begin{equation}
\left\langle \varepsilon_{\mathrm{tot}}\right\rangle =\left\langle \varepsilon_{\mathrm{ord}}\right\rangle +\left\langle \varepsilon_{\mathrm{vac}}\right\rangle .\label{eq:6}\end{equation}
From this we can see that the effective cosmological constant in density
formation would be \begin{equation}
\varepsilon_{\mathrm{eff}}=\frac{\lambda c^{4}}{8\pi G}+\left\langle \varepsilon_{\mathrm{vac}}\right\rangle .\label{eq:7}\end{equation}
Now let us introduce the critical density\begin{equation}
\varepsilon_{\mathrm{crit}}\equiv\frac{3c^{2}H_{0}^{2}}{8\pi G}\simeq\unitfrac[5.16\times10^{3}]{eV}{cm^{3}},\label{eq:8}\end{equation}
with the Hubble constant at its present day value of $H_{0}\approx\unitfrac[70]{\unitfrac{km}{s}}{Mpc}$.
This leaves a value for the effective density as\begin{equation}
\left|\varepsilon_{\mathrm{eff}}\right|\simeq\varepsilon_{\mathrm{crit}}\times73\%\simeq\unitfrac[3.76\times10^{3}]{eV}{cm^{3}}.\label{eq:9}\end{equation}
In addition, the vacuum density can be calculated by summing the zero-point
energies of all normal modes $k$ of some field of mass $m$ up to
a wave cutoff $\Lambda_{\mathrm{cut}}\gg\nicefrac{mc}{\hbar}$, as\begin{equation}
\left\langle \varepsilon_{\mathrm{vac}}\right\rangle =\frac{1}{\left(2\pi\right)^{3}}\int_{0}^{\Lambda_{\mathrm{cut}}}\frac{\hbar}{2}\sqrt{k^{2}c^{2}+\frac{m^{2}c^{4}}{\hbar^{2}}}\cdot4\pi k^{2}dk\simeq\frac{\hbar c\Lambda_{\mathrm{cut}}^{4}}{16\pi^{2}}.\label{eq:10}\end{equation}
Assuming that the smallest limit of general relativity is the Planck
scale, we can take $\Lambda_{\mathrm{cut}}=\pi\sqrt{\nicefrac{c^{3}}{\hbar G}}$
into \prettyref{eq:10} to get

\begin{equation}
\left\langle \varepsilon_{\mathrm{vac}}\right\rangle \simeq\frac{\pi^{3}c^{7}}{16\hbar G^{2}}\simeq\unitfrac[5.60\times10^{126}]{eV}{cm^{3}}.\label{eq:11}\end{equation}
This is much more huge than the effective density as witnessed in
reality. For the real world in which we live, we need Einstein's cosmological
term in order to cancel out the vacuum density of $\left|\left\langle \varepsilon_{\mathrm{vac}}\right\rangle +\nicefrac{\lambda c^{4}}{8\pi G}\right|$
to more than 123 decimal places. It is the famous \textquotedbl{}old
problem\textquotedbl{} of the cosmological constant. On the other
hand, the \textquotedbl{}new problem\textquotedbl{} has arisen \cite{key-1,key-2}
because modern observations give us the very small but nonzero value
of \prettyref{eq:9}.

Further, as outlined in the abstract, there is a large gulf that separates
certain theories. On one side is the theory of the inflationary universe
that deals with the growing scale factor before $\unit[10^{-36}]{s}$
in cosmic time; on the other, is the theory of dark energy that describes
an accelerating expansion universe at about the range of $z<2$. Of
course, this represents a massive difference in cosmic time. Nevertheless,
despite the discrepancy, we still wish to have a complete picture
of our universe. Following this idea, we shall try to use inflationary
theory as the framework for our discussion in this paper.

And now to an overview of our journey: In Section II, I will give
a brief review of inflationary theory - the elegant explanation that
gives us many beautiful solutions to the problems of the big bang
theory. Before discussing our new proposal, it is most instructive
to touch upon this topic. In Section III, I will introduce classical
scalar field (inflaton) dynamics to an universe with a time-varying
inflaton potential in order to find a new solution for the scale factor.
In Section IV, some results of toy models will be presented to provide
a clear image of the proposal and in the final section I will give
a full discussion of the new proposal and try to answer the problems
which have been mentioned above.

An extraneous but important point should be included here: I would
like to dedicate this work to my sweet daughter CoCo, a lovely cat
who was smart, charming and kind. She gave me much joy, support and
inspiration and I wish to thank her for accompanying me during the
past 11 years, especially through the nights when I was working and
studying. During these times, if I couldn\textquoteright{}t sleep,
she didn\textquoteright{}t sleep and it is thanks to her that I was
reminded to recheck my solutions once again, searching for the important
details that I had previously missed. This was her last gift before
she left and it\textquoteright{}s very sad for me: she passed away
on Dec. 21, 2010.

\section{review of the inflationary universe theory}

\subsection*{Motivation}

After the day when Lemaître proposed what would later become known
as the Hot Big Bang theory \cite{key-12}, cosmology transformed into
a famous and precise discipline of physics. Finally, we were able
to explore a reasonable picture of the universe without resorting
to romantic and religious concepts and, consequently, puzzles like
the origin of matter, the age of the universe and other complicated
problems can be solved in the present day. Progress was further complimented
when Gamow et al. \cite{key-13,key-14,key-15} predicted the remnant
temperature that we now call cosmic microwave background radiation
(CMBR) and thereby underlined Lemaître\textquoteright{}s theory as
a compelling explanation for the emergence of our universe. Regardless,
even though we have achieved so much, many unsolved problems remain.
The following is a list of difficulties that arose from the hot big
bang theory and thus brought about the inflationary theory \cite{key-16}:
\begin{enumerate}
\item The homogenous and isotropic problem: according to observations, the
universe is homogenous and isotropic in large scales. What is the
reason for this?
\item The horizon problem: considering the initial length and the causal
length close to the era of the Planck scale, we find a huge value
for the ratio: \begin{equation}
\frac{l_{\mathrm{initial}}}{l_{\mathrm{causal}}}=\frac{\frac{ct_{\mathrm{now}}}{\mathrm{R}_{\mathrm{now}}}\mathrm{R}_{\mathrm{Planck}}}{ct_{\mathrm{Planck}}}\simeq10^{28}.\label{eq:12}\end{equation}
This is dependent on the scale-time-temperature relation \begin{equation}
\mathrm{R}\left(t\right)\varpropto\sqrt{t}\varpropto T^{-1}\left(t\right).\label{eq:13}\end{equation}
\prettyref{eq:12} tells us that the region of CMBR that we see today
is much bigger than the horizon at the last scattering.
\item The flatness problem: according to the 2nd Fridemann equation, but
with an arbitrary curvature parameter $\mathrm{K}$, yields $\Omega\left(t\right)-1=\nicefrac{\mathrm{K}}{\left(H\mathrm{R}\right)^{2}}$.
If we suppose the expansion of the universe is uniform , we find\begin{equation}
\frac{\Omega\left(t_{\mathrm{Planck}}\right)-1}{\Omega\left(t_{\mathrm{now}}\right)-1}\simeq10^{-56}.\label{eq:14}\end{equation}
This shows that the universe should be flat $\left(\mathrm{K}\approx0\right)$
during its very early stage.
\item The initial perturbation problem: perturbation must be $\frac{\delta\varepsilon}{\varepsilon}\sim10^{-5}$
on galactic scales to explain the large-scale structure of the universe.
\item The magnetic-monopole problem: the Grand Unified Theory (GUT) informs
that lots of magnetic monopoles must have been created in the extreme
heat of the early universe \cite{key-17,key-18}. However, we are
yet to find any in the present day. 
\item The total mass problem: the total mass of the observable part of the
universe is $\sim\unit[10^{60}]{M_{Planck}}$.
\item The total entropy problem: the total entropy we observe today is greater
than $10^{87}$.
\end{enumerate}

\subsection*{Inflation as scalar field dynamics}

It is helpful now to mention a brief early history of inflationary
theory. In 1974, Linde was the first to realize that the energy density
of a scalar field plays the role of the vacuum energy/cosmological
constant \cite{key-19}. Then, in 1979 - 1980, Starobinsky wrote the
first semi-realistic model of an inflationary type \cite{key-20}.
Meanwhile, at the end of the 1970s, Guth investigated the magnetic-monopole
problem and found that a positive-energy false vacuum would generate
an exponential expansion of space \cite{key-22}. The idea which he
proposed is the model we call \textquotedbl{}old inflation\textquotedbl{}
today. Unfortunately, it is afflicted by a certain problem: the probability
of bubble formation would cause the universe either to be extremely
inhomogeneous by way of an inflation period that was too short or
to contain a long period of inflation and a separate open universe
with a vanishingly small cosmological parameter $\Omega$ \cite{key-23,key-24,key-25}.
Soon, therefore, a theory called \textquotedbl{}new inflation\textquotedbl{}
was proposed \cite{key-26,key-27}. It suggested a scenario whereby
the inflaton field $\phi$ should slowly roll down to the minimum
of its effective potential. During slow-roll inflation, energy is
released homogeneously into the whole of space and density perturbations
are inversely proportional to $\dot{\phi}$ \cite{key-28,key-29,key-30,key-31,key-32,key-33}.

Following the brief but incomplete review above, we will now turn
our attention to the construction of basic inflationary theory. Consider
the action of our universe without the cosmological constant in Planck
units, $c=G=\hbar=1$, \begin{equation}
S_{\mathrm{u}}=\frac{1}{16\pi}\int d^{4}x\sqrt{-g}R+\int d^{4}x\sqrt{-g}\mathcal{L}_{\mathrm{m}},\label{eq:15}\end{equation}
where $R$ is Ricci scalar and $g$ is the determinate of a spacetime
metric tensor. Due to the fact that inflation began before the GUT
phase transition, we could say that the Lagrangian of matter was made
by a dimensionless scalar field $\phi\left(x^{\mu}\right)$, \begin{equation}
\mathcal{L}_{\mathrm{m}}\left(\phi,\partial_{\mu}\phi,x^{\mu}\right)=\frac{1}{2}g^{\mu\nu}\partial_{\mu}\phi\partial_{\nu}\phi-V\left(\phi,x^{\mu}\right).\label{eq:16}\end{equation}
When we vary \prettyref{eq:15} to $g_{\mu\nu}$ by the variation
principle, we get the Einstein field equation\begin{equation}
R_{\mu\nu}-\frac{1}{2}Rg_{\mu\nu}=-16\pi\left(\frac{\delta\mathcal{L}_{\mathrm{m}}}{\delta g^{\mu\nu}}-\frac{1}{2}g_{\mu\nu}\mathcal{L}_{\mathrm{m}}\right).\label{eq:17}\end{equation}

Look at the two equations \prettyref{eq:16} and \prettyref{eq:17}.
There are two keys to these equations that would enable us to investigate
the universe: one is to give the structure of spacetime, i.e. the
metric tensor {}``$g_{\mu\nu}$''; the other is to suggest a model
of the matter field, i.e. the potential term {}``$V\left(\phi,x^{\mu}\right)$''.
However, \prettyref{eq:17} tells us that the situation is too complex
as $g_{\mu\nu}$ and $\phi$ vigorously interact with each other.
To simplify, let us consider the formula in the bracket of \prettyref{eq:17}:
the energy-momentum tensor. Another formation in scalar field is\begin{equation}
T_{\mu\nu}=\partial_{\mu}\phi\partial_{\nu}\phi-g_{\mu\nu}\left[\frac{1}{2}g^{\alpha\beta}\partial_{\alpha}\phi\partial_{\beta}\phi-V\left(\phi,x^{\mu}\right)\right].\label{eq:18}\end{equation}
According to observations, the spacetime metric tensor should be off-diagonal
as much as possible. For this reason, we want $T_{\mu\nu}$ to approach
the off-diagonal as well. When we attempt to separate the scalar field
into two parts\begin{equation}
\phi\left(t,x^{i}\right)=\phi\left(t\right)+\delta\phi\left(t,x^{i}\right),\label{eq:19}\end{equation}
we find that the amplitude of $\delta\phi\left(t,x^{i}\right)$ must
be small enough to make the tensors adhere to the off-diagonals that
we desire. 

In passing through the above discussion, we become confident that
$\phi\left(t\right)$ has a major role in affecting spacetime geometry.
Therefore, the line element of the Friedmann\textendash{}Lemaître\textendash{}Robertson\textendash{}Walker
(FLRW) spacetime can be introduced here as the spacetime background
\begin{equation}
ds^{2}=dt^{2}-\mathrm{R}^{2}\left(t\right)\left[\frac{dr^{2}}{1-\mathrm{K}r^{2}}+r^{2}d\theta^{2}+r^{2}\sin^{2}\theta d\varphi^{2}\right],\label{eq:20}\end{equation}
where $\mathrm{R}\left(t\right)$ is the scale factor and $\mathrm{K}$
is the curvature parameter. Now taking \prettyref{eq:20} into \prettyref{eq:17}
with the time dependent scalar field $\phi\left(t\right)$ and making
the general consideration that the potential function of \prettyref{eq:16}
is only dependent on $\phi$, we obtain the Fridemann equations corresponding
to the scalar field:\begin{equation}
\mathrm{\frac{\ddot{R}}{R}}=-\frac{8\pi}{3}\left(\dot{\phi}^{2}-V\left(\phi\right)\right),\label{eq:21}\end{equation}
\begin{equation}
\left(\mathrm{\frac{\dot{R}}{R}}\right)^{2}=\frac{8\pi}{3}\left(\frac{1}{2}\dot{\phi}^{2}+V\left(\phi\right)\right)-\frac{\mathrm{K}}{\mathrm{R^{2}}}.\label{eq:22}\end{equation}
Furthermore, from the fact that energy-momentum conservation requires\begin{equation}
D_{\mu}T^{\mu\nu}=0,\label{eq:23}\end{equation}
where the operator $D_{\mu}$ is the covariant derivative, we obtain
the scalar field equation \begin{equation}
\ddot{\phi}+3\left(\mathrm{\frac{\dot{R}}{R}}\right)\dot{\phi}+\frac{dV\left(\phi\right)}{d\phi}=0\label{eq:24}\end{equation}
by taking \prettyref{eq:18} into \prettyref{eq:23}.

To find the solution for the scale factor during inflation, two conditions
should be noted:
\begin{enumerate}
\item $\dot{\phi}^{2}\ll V\left(\phi\right)$ initially makes $\ddot{\mathrm{R}}\gg0$.
\item To avoid the bubble-formation problem, the slow-roll scenario requires
$\ddot{\phi}\approx0$ during the period of inflation.
\end{enumerate}
Given the above two conditions and neglecting the curvature term $\nicefrac{\mathrm{K}}{\mathrm{R^{2}}}$
in \prettyref{eq:22} (actually, even if we keep this term to begin
with, the initial stages of inflation will soon render it obsolete),
potential models for inflation must satisfy the following:
\begin{enumerate}
\item By calculating the approximation of $\nicefrac{-\dot{H}}{H^{2}}$,
we have two slow-roll parameters as defined by Liddle and Lyth \cite{key-34}
\begin{equation}
\epsilon\left(\phi\right)\equiv-\frac{\dot{H}}{H^{2}}\simeq\frac{1}{16\pi}\left(\frac{V^{'}\left(\phi\right)}{V\left(\phi\right)}\right)^{2}\ll1,\label{eq:25}\end{equation}
\begin{equation}
\eta\left(\phi\right)\equiv\epsilon\left(\phi\right)-\frac{\ddot{\phi}}{H\dot{\phi}}\simeq\frac{1}{8\pi}\frac{V^{''}\left(\phi\right)}{V\left(\phi\right)}\ll1.\label{eq:26}\end{equation}

\item According to the horizon problem, \prettyref{eq:12} and \prettyref{eq:13}
tell us $\nicefrac{\mathrm{R}_{\mathrm{now}}}{\mathrm{R}_{\mathrm{Planck}}}\simeq10^{28}\approx e^{65}$.
Therefore, the e-folding number $N$ should be\begin{equation}
N\equiv\ln\frac{\mathrm{R}\left(t_{e}\right)}{\mathrm{R}\left(t_{i}\right)}\simeq-8\pi\int_{\phi_{i}}^{\phi_{e}}\left(\frac{V\left(\phi\right)}{V^{'}\left(\phi\right)}\right)d\phi\gtrsim60-70,\label{eq:27}\end{equation}
 where the suffix $i$ means the beginning of inflation and the suffix
$e$ means the end of inflation.
\end{enumerate}

\section{the ghost of inflaton}

Generally, people introduce models of $V\left(\phi\right)$ for the
inflation corresponding to the above discussion and, through this
method, much success can be achieved. However, the setting of $\ddot{\phi}\approx0$
during this inflation means that contributions to the damping term
$-3H\dot{\phi}$ are received from the potential energy difference
alone and in entirety. The setting also means that the contribution
of $\ddot{\phi}$ to the damping term is prevented, and the energy
exchange between the potential and kinetic terms is also turned off.
Therefore, the method not only disqualifies us from obtaining a solution
for the scale factor after inflation (because the scenario is specific
to our universe \emph{during} inflation), but also limits study to
a special case for the three cosmic field equations (\prettyref{eq:21},
\prettyref{eq:22} and \prettyref{eq:24}). In my opinion, even if
we only have an interest in our universe, we do not need to concern
ourselves with the assumption $\ddot{\phi}\approx0$ during inflation,
providing that we already know the proper inflaton models. Therefore,
let us try to consider another assumption: First, we allow that the
potential term of the Lagrangian \prettyref{eq:16} is time-varied
as $V\left(\phi,t\right)$. Then, we can easily obtain the new cosmic
field equations

\begin{equation}
\mathrm{\frac{\ddot{R}}{R}}=-\frac{8\pi}{3}\left(\dot{\phi}^{2}-V\left(\phi,t\right)\right),\label{eq:28}\end{equation}
\begin{equation}
\left(\mathrm{\frac{\dot{R}}{R}}\right)^{2}=\frac{8\pi}{3}\left(\frac{1}{2}\dot{\phi}^{2}+V\left(\phi,t\right)\right)-\frac{\mathrm{K}}{\mathrm{R^{2}}},\label{eq:29}\end{equation}

\begin{equation}
\ddot{\phi}\dot{\phi}+3\left(\mathrm{\frac{\dot{R}}{R}}\right)\dot{\phi}^{2}+\frac{dV\left(\phi,t\right)}{dt}=0\label{eq:30}\end{equation}
by calculating \prettyref{eq:17} and \prettyref{eq:23} with the
FLRW spacetime background \prettyref{eq:20}. Consequently, the equation\begin{equation}
-3H\dot{\phi}^{2}=\frac{d}{dt}\left(\frac{1}{2}\dot{\phi}^{2}+V\left(\phi,t\right)\right)\label{eq:31}\end{equation}
gained from \prettyref{eq:30} is the screen for our journey, whereupon
we can obtain the scale factor solution \begin{equation}
\mathrm{R}\left(t\right)=\mathrm{R}\left(t_{i}\right)\exp\left[\int_{t_{i}}^{t}\left(H\left(t_{i}\right)-4\pi\int_{t_{i}}^{T}\dot{\phi}^{2}d\tau\right)dT\right]\label{eq:32}\end{equation}
by taking \prettyref{eq:29} (neglecting the curvature term $\nicefrac{\mathrm{K}}{\mathrm{R^{2}}}$)
into \prettyref{eq:31}. $t_{i}$ denotes the cosmic time when inflation
was beginning; $t$ is an arbitrary cosmic time after $t_{i}$; $H\left(t_{i}\right)$
is a constant that needs to be determinate, called the initial Hubble
parameter (IHP); and $\dot{\phi}^{2}$ is the kinetic energy term
of inflaton with a value that is never negative. Of course, $H\left(t\right)=0$
is another trivial solution of \prettyref{eq:31} which does not require
our concern. Next, taking the first and second derivative of $\mathrm{R}\left(t\right)$,
we have\begin{equation}
\left(\frac{\mathrm{\dot{R}}\left(t\right)}{\mathrm{R}\left(t\right)}\right)^{2}=\left[H\left(t_{i}\right)-4\pi\int_{t_{i}}^{t}\dot{\phi}^{2}d\tau\right]^{2},\label{eq:33}\end{equation}
\begin{equation}
\frac{\ddot{\mathrm{R}}\left(t\right)}{\mathrm{R}\left(t\right)}=\left[H\left(t_{i}\right)-4\pi\int_{t_{i}}^{t}\dot{\phi}^{2}d\tau\right]^{2}-4\pi\dot{\phi}^{2}\left(t\right).\label{eq:34}\end{equation}
Now, by incorporating \prettyref{eq:34} into \prettyref{eq:28},
we can rebuild the time-dependent potential as \begin{equation}
V\left(t\right)=\frac{3}{8\pi}\left[H\left(t_{i}\right)-4\pi\int_{t_{i}}^{t}\dot{\phi}^{2}d\tau\right]^{2}-\frac{1}{2}\dot{\phi}^{2}\left(t\right).\label{eq:35}\end{equation}
Thus, we can define the Hubble-$\lambda$-function that appears in
\prettyref{eq:32}, \prettyref{eq:33}, \prettyref{eq:34} and \prettyref{eq:35}
as\begin{equation}
H_{\lambda}\left(t\right)=H\left(t_{i}\right)-4\pi\int_{t_{i}}^{t}\dot{\phi}^{2}d\tau.\label{eq:36}\end{equation}
To observe \prettyref{eq:36}, the integral should be\begin{equation}
\int_{t_{i}}^{t>t_{r}}\dot{\phi}^{2}d\tau=\int_{t_{i}}^{t_{r}}\dot{\phi}^{2}d\tau+\int_{t_{r}}^{t>t_{r}}\dot{\phi}^{2}d\tau=\int_{t_{i}}^{t_{r}}\dot{\phi}^{2}d\tau\label{eq:37}\end{equation}
if $\dot{\phi}$ is at rest after time $t_{r}$: i.e. $\dot{\phi}\left(t\geq t_{r}\right)=0$.
Therefore, $\left(\nicefrac{\dot{\mathrm{R}}}{\mathrm{R}}\right)_{t\geq t_{r}}^{2}$
and $\left(\nicefrac{\ddot{\mathrm{R}}}{\mathrm{R}}\right)_{t\geq t_{r}}$
are non-negative constants. Comparing the 2nd Friedmann equation which
is only dependent on the cosmological constant, \begin{equation}
\left(\mathrm{\frac{\dot{R}\left(t\right)}{R\left(t\right)}}\right)^{2}=\frac{\Lambda}{3},\label{eq:38}\end{equation}
with \prettyref{eq:33}, we come to \begin{equation}
\Lambda=3H_{\lambda}^{2}\left(t\geq t_{r}\right)=3\left[H\left(t_{i}\right)-4\pi\int_{t_{i}}^{t_{r}}\dot{\phi}^{2}d\tau\right]^{2}.\label{eq:39}\end{equation}
Moreover, reviewing the earlier discussion of \prettyref{eq:7}, we
can treat \begin{equation}
\left\langle \varepsilon_{\mathrm{vac}}\right\rangle =\frac{3H^{2}\left(t_{i}\right)}{8\pi}\label{eq:40}\end{equation}
as the vacuum energy density. Then the term for Einstein\textquoteright{}s
cosmological term would be \begin{equation}
\lambda=-24\pi H\left(t_{i}\right)\int_{t_{i}}^{t_{r}}\dot{\phi}^{2}d\tau+48\pi^{2}\left(\int_{t_{i}}^{t_{r}}\dot{\phi}^{2}d\tau\right)^{2}.\label{eq:41}\end{equation}
Therefore, \prettyref{eq:39} is the effective cosmological constant
that we know of phenomenally. Meanwhile, the potential $V\left(t\geq t_{r}\right)$
will also land on a fixed positive value $\nicefrac{\Lambda}{8\pi}$,
if -- and only if -- $t_{r}$ exists. This is why we view the cosmological
constant as a ghost of inflaton. 

Clearer solution-behavior can be seen in the following table. For
this to be fully comprehensible, it should be noted that $t_{0}$
is the characteristic time when $H_{\lambda}\left(t_{0}\right)=0$;
$t_{r}$ is the time when $\dot{\phi}$ begins to be at rest; $\dot{\phi}=0$
does not mean that the inflaton always stops - it merely expresses
something like the speed of an oscillator at its turning point; $\Lambda_{i}$
is the effective cosmological constant of a type $i$ universe; and
the time dependent variable is denoted by $v\left(t\right)$ (we use
\textquotedbl{}$\searrow$\textquotedbl{} to describe its decrease).

\begin{table}[H]
\centering{}\begin{tabular}{>{\raggedright}m{1.3cm}>{\raggedright}m{4.6cm}>{\raggedright}m{2.1cm}>{\raggedright}m{2.1cm}>{\raggedright}m{3.1cm}>{\raggedright}m{2cm}}
\hline 
 & Type 1 & Type 2 & Type 3 & Type 4 & Type 5\tabularnewline
 & $t>t_{0}$; without $t_{r}$ & $t\geq t_{r}>t_{0}$ & $t\geq t_{0}=t_{r}$ & $t\geq t_{r}$; without $t_{0}$ & $t\ll t_{0}$, $t_{r}$\tabularnewline
\hline 
$\dot{\phi}^{2}\left(t\right)$ & $\geq0$ & at rest & at rest & at rest & $\geq0$\tabularnewline
$H_{\lambda}\left(t\right)$ & $-\left\Vert v\left(t\right)\right\Vert $, $\searrow$ & $-\sqrt{\frac{\Lambda_{2}}{3}}$ & 0 & $\sqrt{\frac{\Lambda_{4}}{3}}$ & $\left\Vert v\left(t\right)\right\Vert $, $\searrow$\tabularnewline
$\frac{\ddot{\mathrm{R}}\left(t\right)}{\mathrm{R}\left(t\right)}$ & $\left(\pm\left\Vert v\left(t\right)\right\Vert \right)\rightarrow\left(+\left\Vert v\left(t\right)\right\Vert \right)$ & $\frac{\Lambda_{2}}{3}$ & 0 & $\frac{\Lambda_{4}}{3}$ & ?\tabularnewline
$V\left(t\right)$ & $\left(\pm\left\Vert v\left(t\right)\right\Vert \right)\rightarrow\left(+\left\Vert v\left(t\right)\right\Vert \right)$ & $\frac{\Lambda_{2}}{8\pi}$ & 0 & $\frac{\Lambda_{4}}{8\pi}$ & ?\tabularnewline
\hline
\end{tabular}\caption{Five types of the evolution of $\mathrm{R}\left(t\right)$. }
\end{table}

As illustrated by the table above, solutions for $\mathrm{R}\left(t\right)$
can be sorted into five types, all of which describe the evolution
of the universe through the existence of characteristic time $t_{0}$
and $t_{r}$. Roughly speaking, the existence of $t_{0}$ is an important
key for demarcating the denouement of the universe. For example, in
a type 1 or type 2 universe, the scale factor would shrink and never
expand again when $4\pi\int_{t_{i}}^{t>t_{0}}\dot{\phi}^{2}d\tau>H\left(t_{i}\right)$
happens. However, if the magnitude of $\dot{\phi}$ comes to rest
quickly enough for there to be no $t_{0}$ existence, as in type 4,
$H_{\lambda}\left(t\geq t_{r}\right)$ will be a positive constant
$\sqrt{\nicefrac{\Lambda_{4}}{3}}$ and the final result a de Sitter
universe. Moreover, a type 3 universe that only has scalar field matter
will be static in the end. Contrastingly, a universe of type 5 is
expanded, but with situations that can not be determined. It is for
this reason that a question mark is introduced to show the uncertainty
of $\nicefrac{\ddot{\mathrm{R}}}{\mathrm{R}}$ and $V\left(t\right)$.

\section{numerical tests}

In this section, we will provide three toy models as tests for our
proposal. Before we can begin, however, the following necessary settings
must be provided: the beginning of time is $t_{i}=0$; the initial
amplitude of the inflaton is $\phi\left(t_{i}\right)=\sqrt{10}$;
the initial size of the scale factor is $\mathrm{R}\left(t_{i}\right)=1$;
the decay parameter of the inflaton is $\beta=\unit[3\pi]{\zeta}$;
and the inflaton mass is $m=\unit[1]{M}$, where the unit of time
is $\unit{\zeta}$ and mass is $\unit{M}$. Attention should be drawn
to the fact that there must be a definition of $\unit{M=\zeta^{-1}}$
in order to satisfy consistency for the units of \prettyref{eq:31}
under the settings of a dimensionless scale factor and inflaton. A
more thorough discussion of units and data analysis will be contained
in the next section. For convenience, the time-varying cosmological
term is defined by $\Lambda\left(t\right)=3H_{\lambda}^{2}\left(t\right)$
and arrows are used to illustrate the evolutionary direction of $V\left(\phi\right)$
in  the following figures.

\subsection{$\boldsymbol{\phi\left(t\right)=\phi\left(t_{i}\right)-\frac{m}{\sqrt{12\pi}}\left(t-t_{i}\right)}$}

This model is the solution to the famous theory of chaotic inflation,
$V\left(\phi\right)=\frac{1}{2}m^{2}\phi^{2}$, during the period
of slow-roll inflation. However, it continues for much longer than
its inflationary period. We can choose the IHP as $H\left(t_{i}\right)=\unit[6.481]{\zeta^{-1}}$
and obtain the following results:

\begin{figure}[H]
\centering{}\includegraphics[scale=0.83]{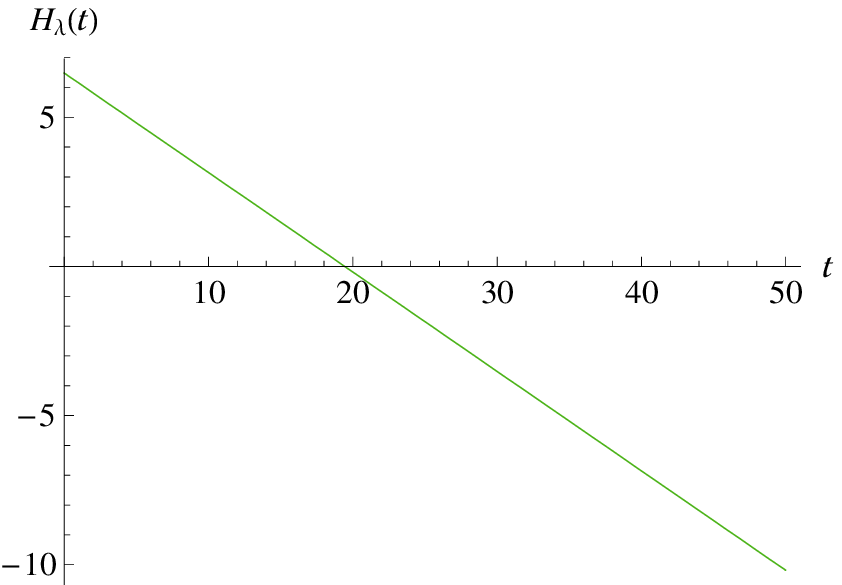}\begin{tabular}{cc}
 & \tabularnewline
\end{tabular}\includegraphics[scale=0.83]{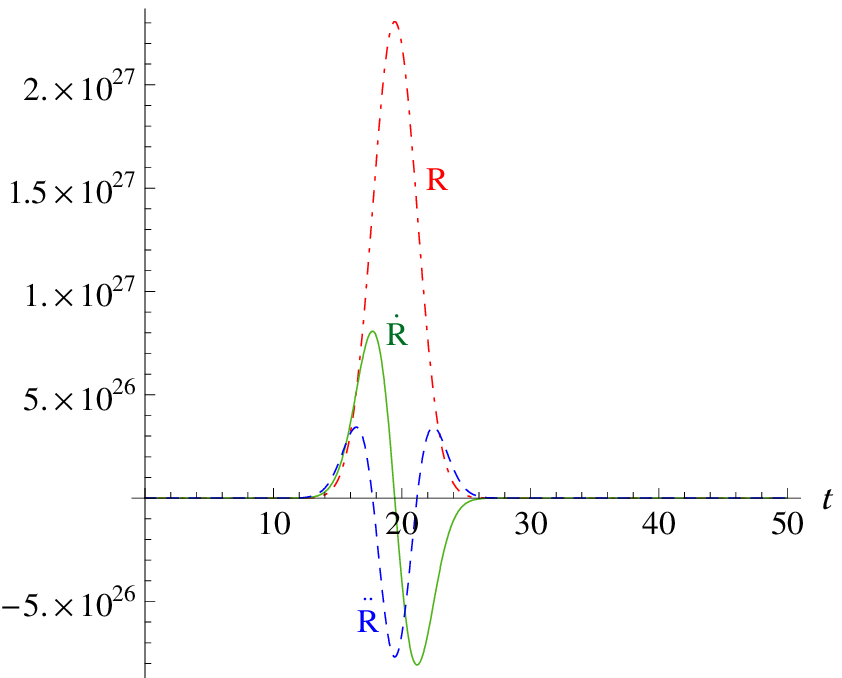}\caption{According to the left picture, we discover both that $H_{\lambda}\left(\unit[19.4430]{\zeta}\right)=0$
(i.e. $t_{0}=\unit[19.4430]{\zeta}$) and $\dot{\mathrm{R}}$ becomes
negative immediately. The negative acceleration occurs during $\unit[17.7109]{\zeta}<t<\unit[21.1751]{\zeta}$.
However, it eventually becomes positive in order to slow down scale
factor shrinkage. Besides, it should be noted that the right picture
only shows amounts and relationship for $\mathrm{R}$, $\dot{\mathrm{R}}$
and $\ddot{\mathrm{R}}$.}
\end{figure}

\begin{figure}[H]
\centering{}\includegraphics[scale=0.83]{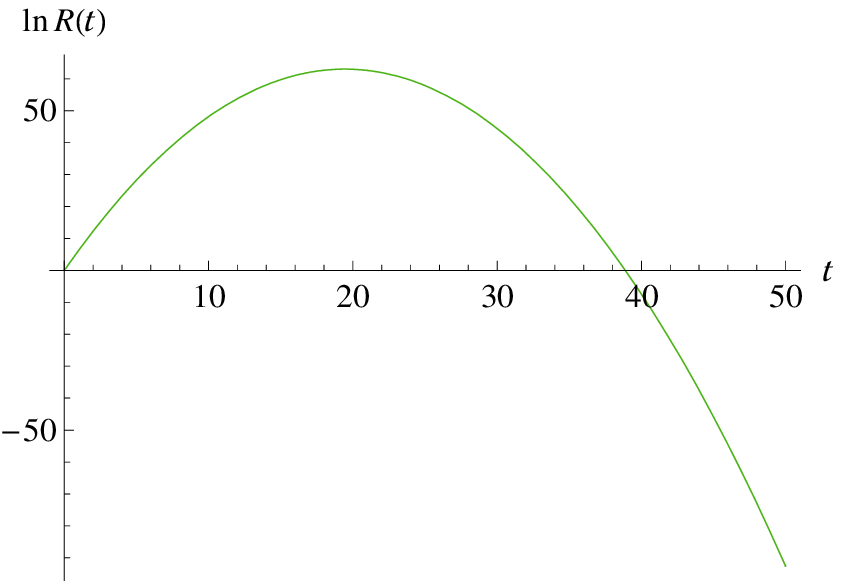}\begin{tabular}{cc}
 & \tabularnewline
\end{tabular}\includegraphics[scale=0.83]{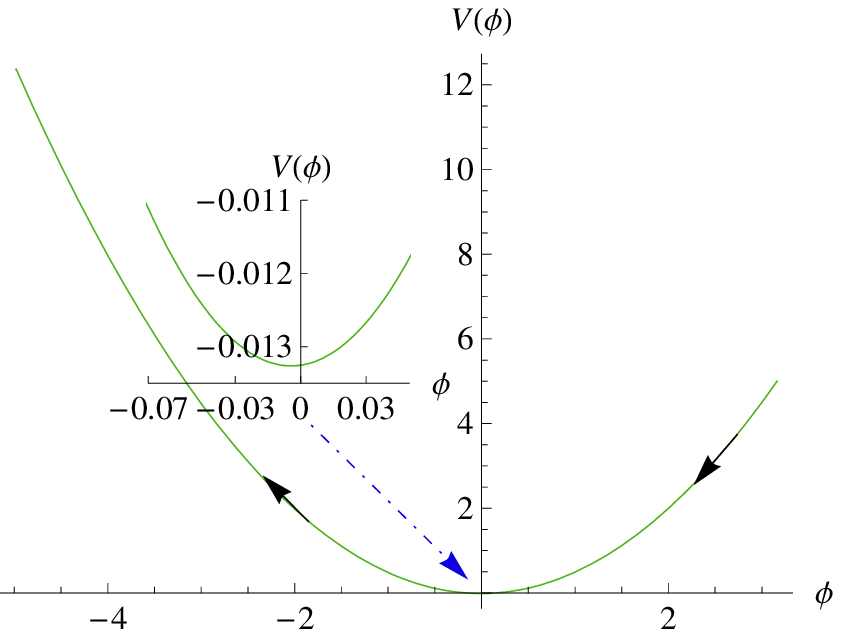}\caption{The scale factor shrinks to one (its initial size) again at $t=\unit[38.886]{\zeta}$.
The time interval of both pictures is $\left[\unit[0]{\zeta},\unit[50]{\zeta}\right]$.
Actually, as the sub-picture shows, the potential would be negative
at $\phi\thickapprox0$. It is almost the same period when negative
acceleration occurs.}
\end{figure}

\subsection{$\boldsymbol{\phi\left(t\right)=\phi\left(t_{i}\right)\exp\left(-\frac{t-t_{i}}{\beta}\right)\cos\left[\frac{m}{\sqrt{12\pi}}\left(t-t_{i}\right)\right]}$}

In this example, we set the IHP as $H\left(t_{i}\right)=\unit[14.523]{\zeta^{-1}}$
and discover an universe in which re-accelerated expansion will take
place after the end of inflation. The results are as follows:

\begin{figure}[H]
\centering{}\includegraphics[scale=0.83]{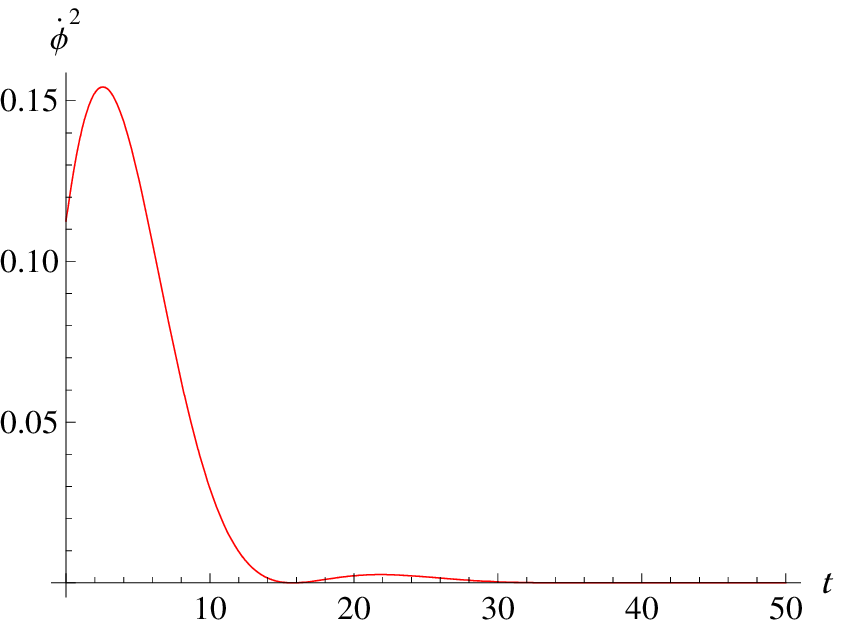}\begin{tabular}{cc}
 & \tabularnewline
\end{tabular}\includegraphics[scale=0.83]{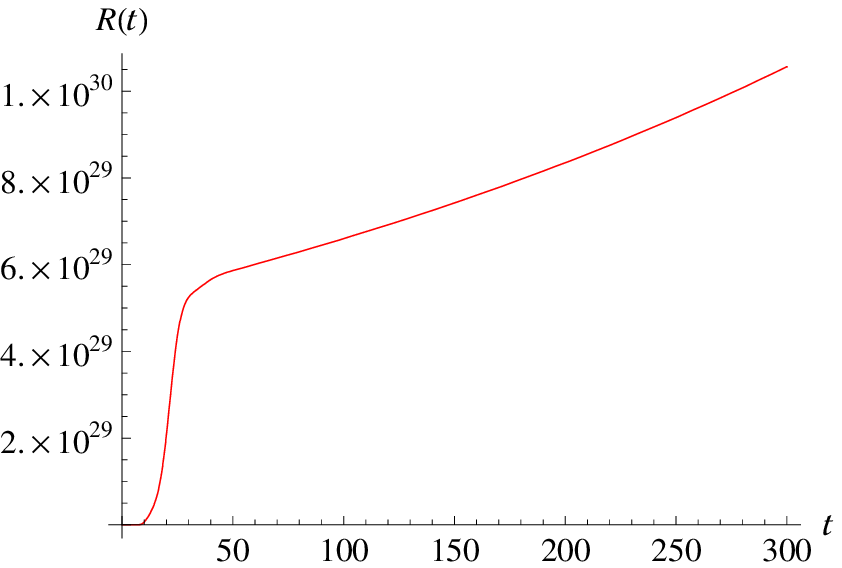}\caption{}
\end{figure}

\begin{figure}[H]
\centering{}\includegraphics[scale=0.83]{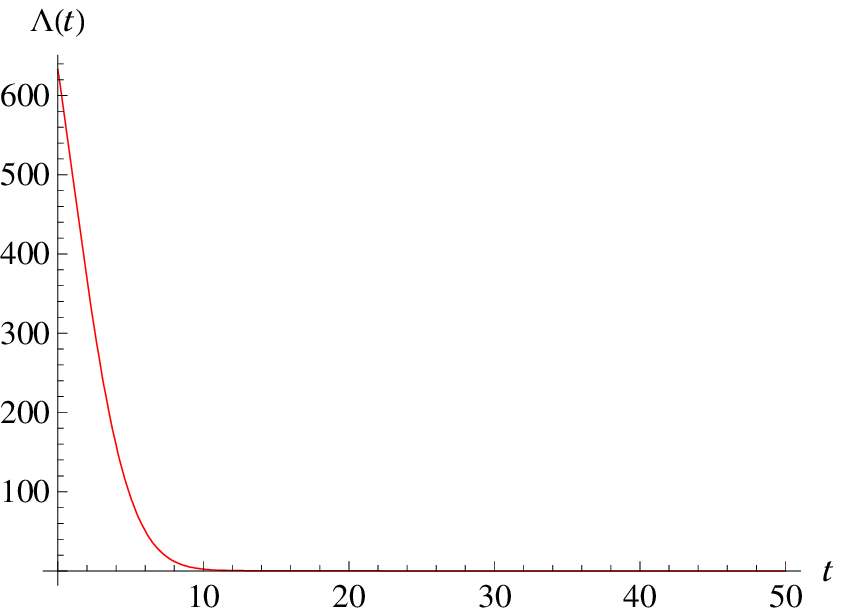}\begin{tabular}{cc}
 & \tabularnewline
\end{tabular}\includegraphics[scale=0.83]{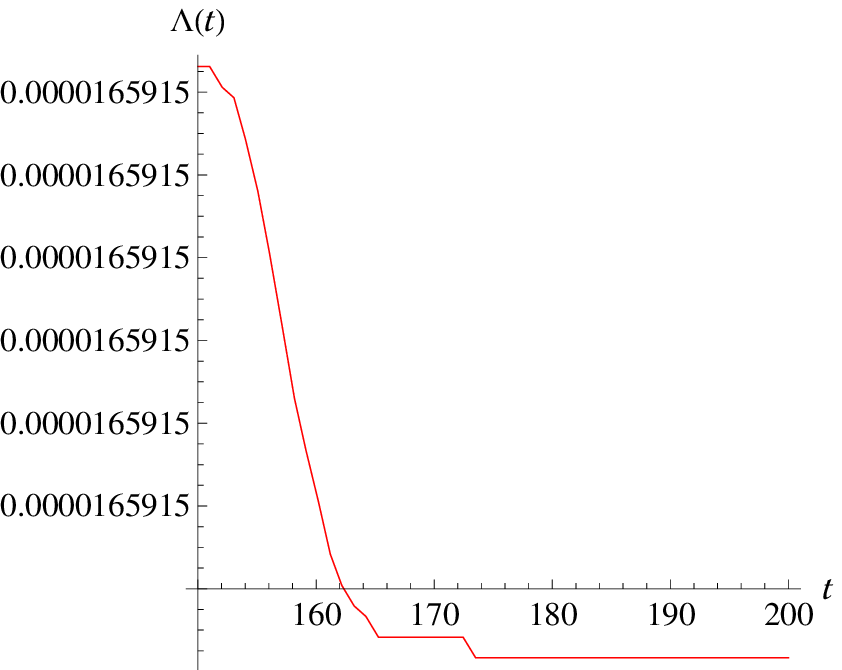}\caption{The cosmological term would become a non-zero positive constant of
less than $\unit[1.6592\times10^{-5}]{\zeta^{-2}}$ after $t\approx\unit[173.4153]{\zeta}$.}
\end{figure}

\begin{figure}[H]
\centering{}\includegraphics[scale=0.83]{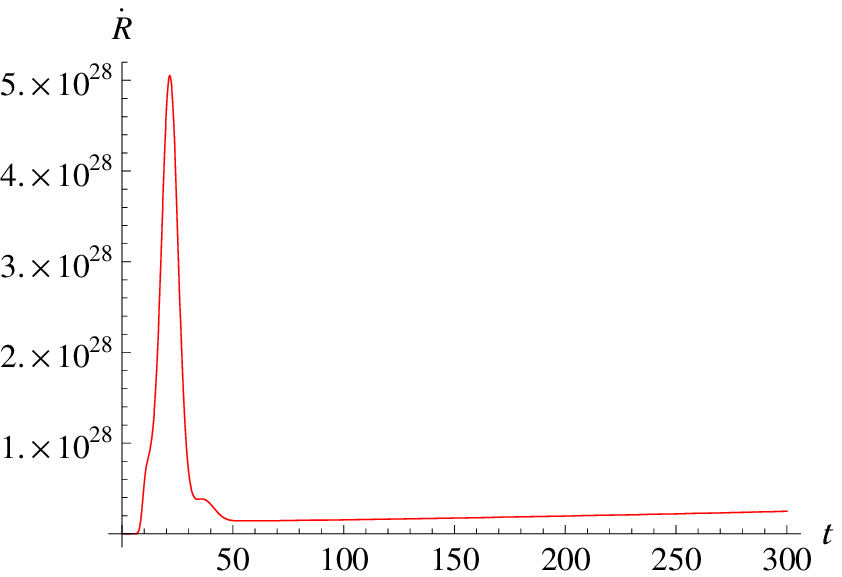}\begin{tabular}{cc}
 & \tabularnewline
\end{tabular}\includegraphics[scale=0.83]{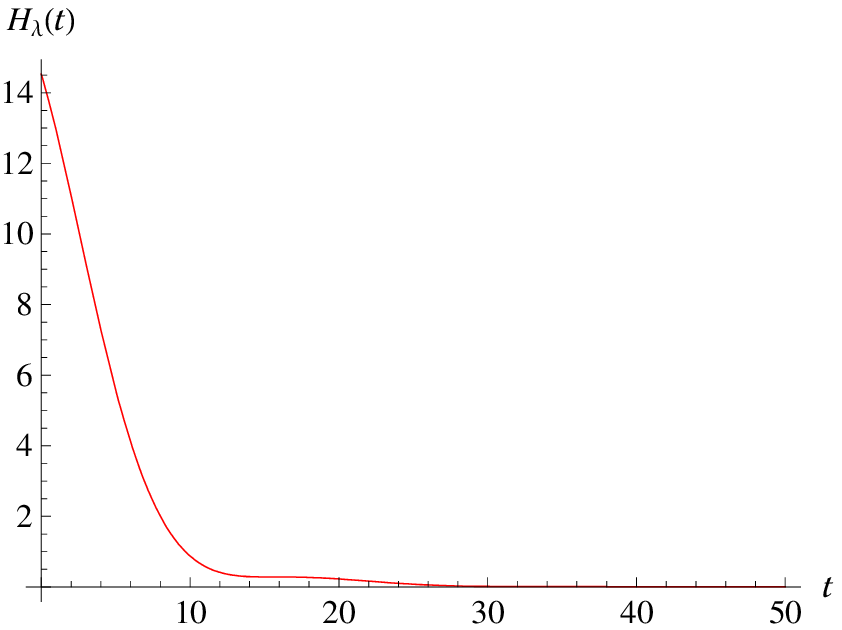}\caption{The end of inflation is approximately $t\thickapprox\unit[21.5302]{\zeta}$
when the maximum of $\dot{\mathrm{R}}$ happens.}
\end{figure}

\begin{figure}[H]
\begin{centering}
\includegraphics[scale=0.83]{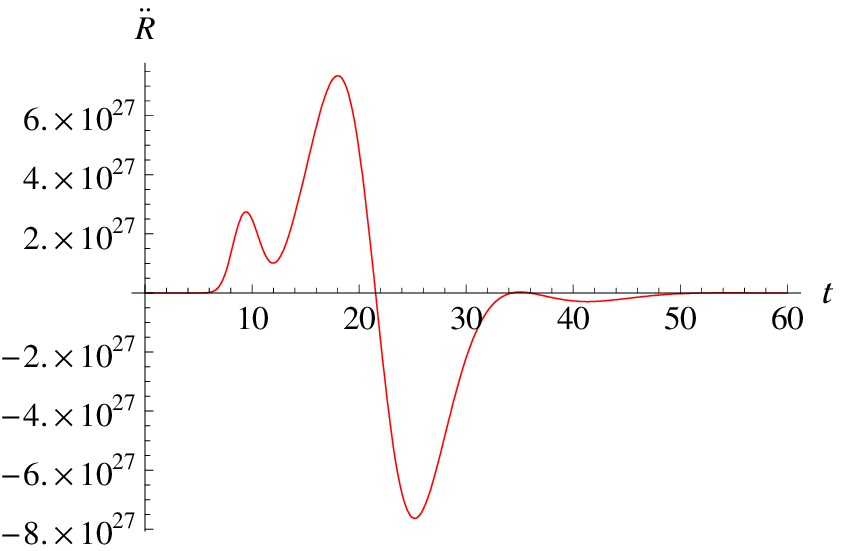}\begin{tabular}{cc}
 & \tabularnewline
\end{tabular}\includegraphics[scale=0.83]{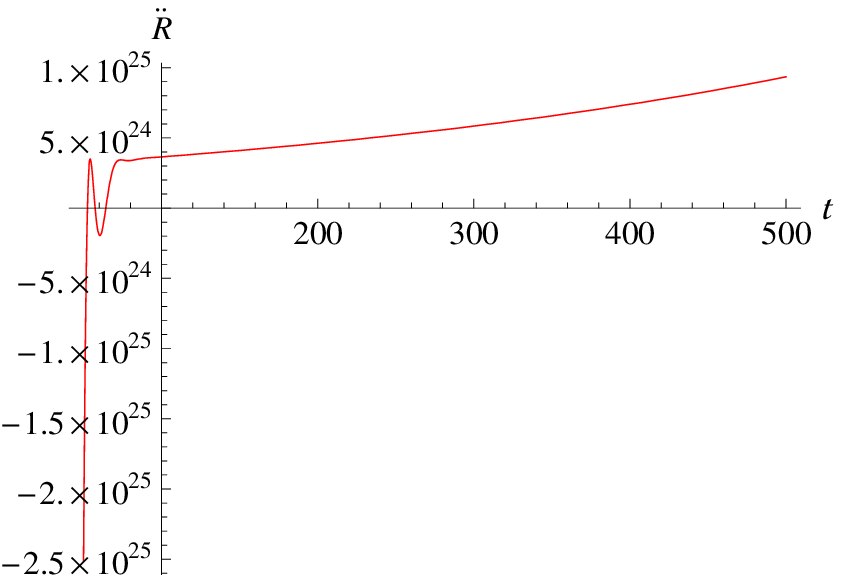}
\par\end{centering}

\caption{$\ddot{\mathrm{R}}$ is no longer less than zero after $t\thickapprox\unit[64.3050]{\zeta}$.}
\end{figure}

\begin{figure}[H]
\centering{}\includegraphics[scale=0.83]{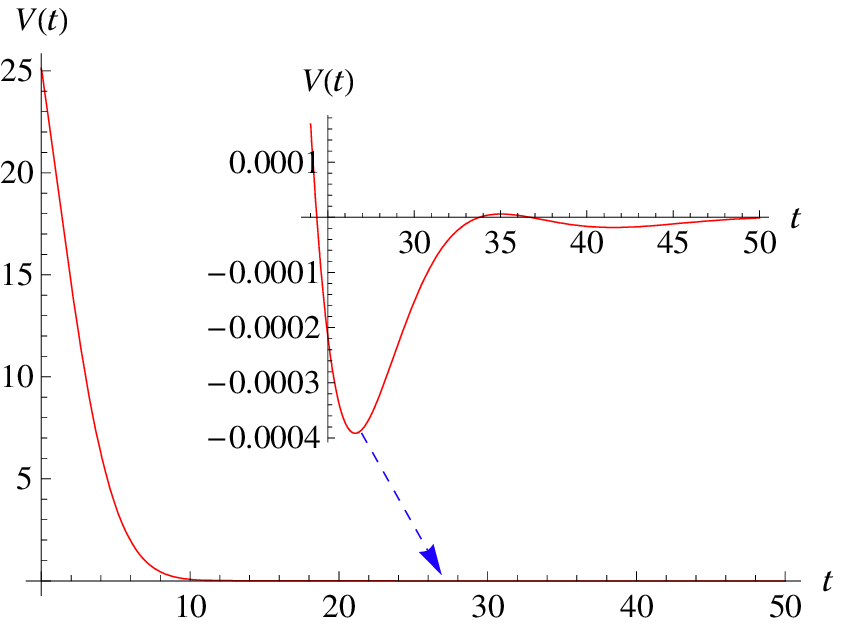}\begin{tabular}{cc}
 & \tabularnewline
\end{tabular}\includegraphics[scale=0.83]{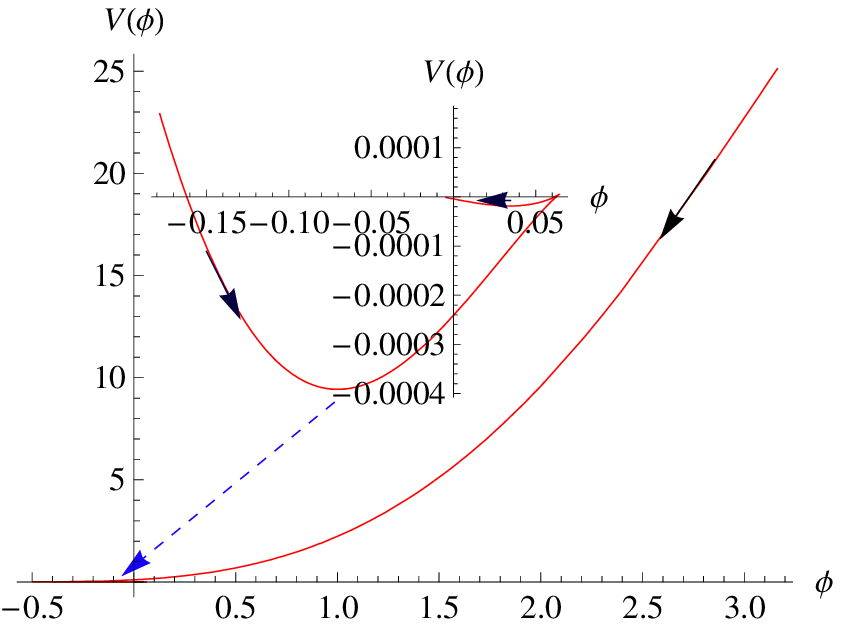}\caption{The time interval of both pictures is $\left[\unit[0]{\zeta},\unit[50]{\zeta}\right]$.
The sub-pictures show the conditions of the minimum negative potential.}
\end{figure}

\begin{figure}[H]
\centering{}\includegraphics[scale=0.83]{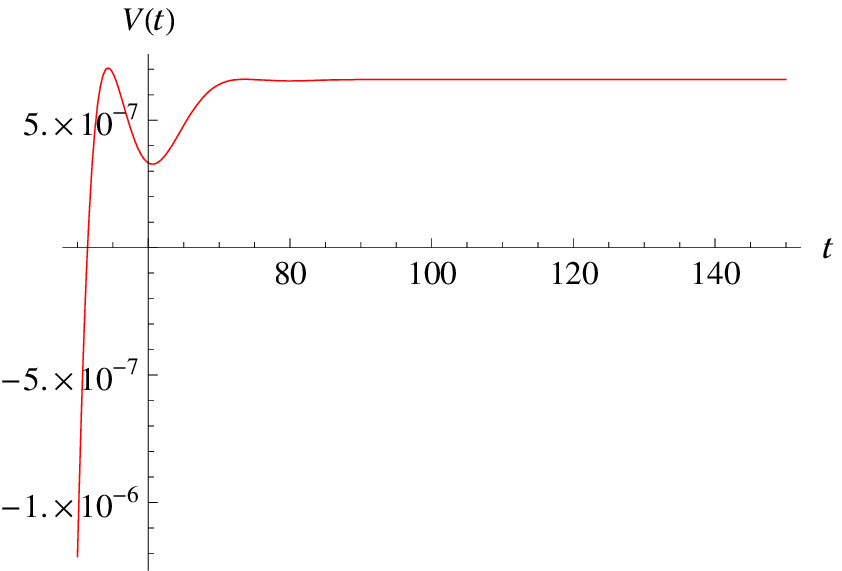}\begin{tabular}{cc}
 & \tabularnewline
\end{tabular}\includegraphics[scale=0.83]{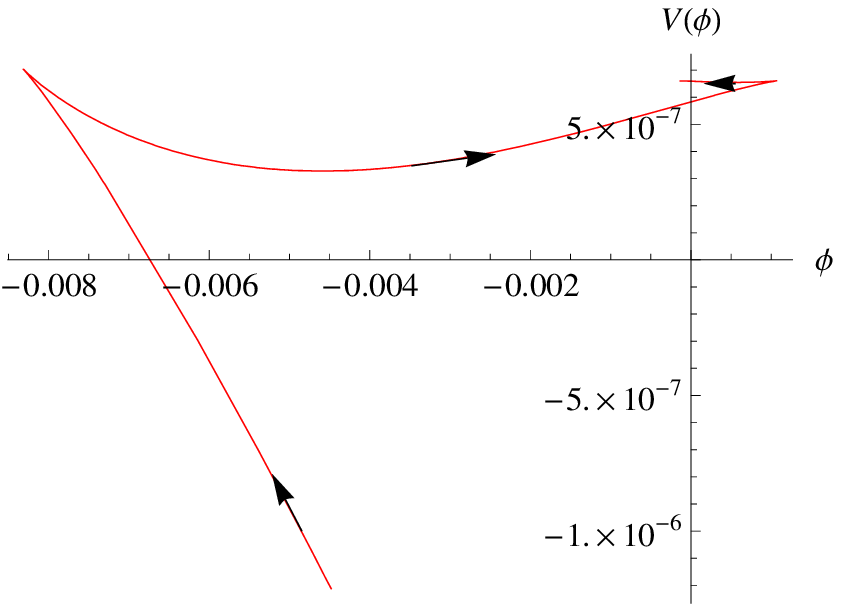}\caption{The time interval of both pictures is $\left[\unit[50]{\zeta},\unit[150]{\zeta}\right]$.
From them, we discover that the potential will rise to positive from
negative, and hold a positive value after a sufficient time has passed.}
\end{figure}

For emphasis, the important data from this example should be mentioned
again: the period of the inflation is before $\unit[21.530]{\zeta}$;
the universe has three instances of negative acceleration during $\unit[21.530]{\zeta}<t<\unit[64.305]{\zeta}$
in order to stop inflation before it immediately emerges into accelerated
expansion again. In this situation, the potential would be $\unit[6.60154\times10^{-7}]{\zeta^{-2}}$
when $t\gg\unit[225]{\zeta}$.

\subsection{$\boldsymbol{\phi\left(t\right)=\phi\left(t_{i}\right)\exp\left(-\frac{t-t_{i}}{\beta}\right)\cos\left[\frac{m}{\sqrt{12\pi}}\left(t-t_{i}\right)\right]}$}

Using the same model as in example B but with the smaller initial
parameter of $H\left(t_{i}\right)=\unit[14.518]{\zeta^{-1}}$, we
find that the scale factor shrinks after the end of inflation.

\begin{figure}[H]
\centering{}\includegraphics[scale=0.83]{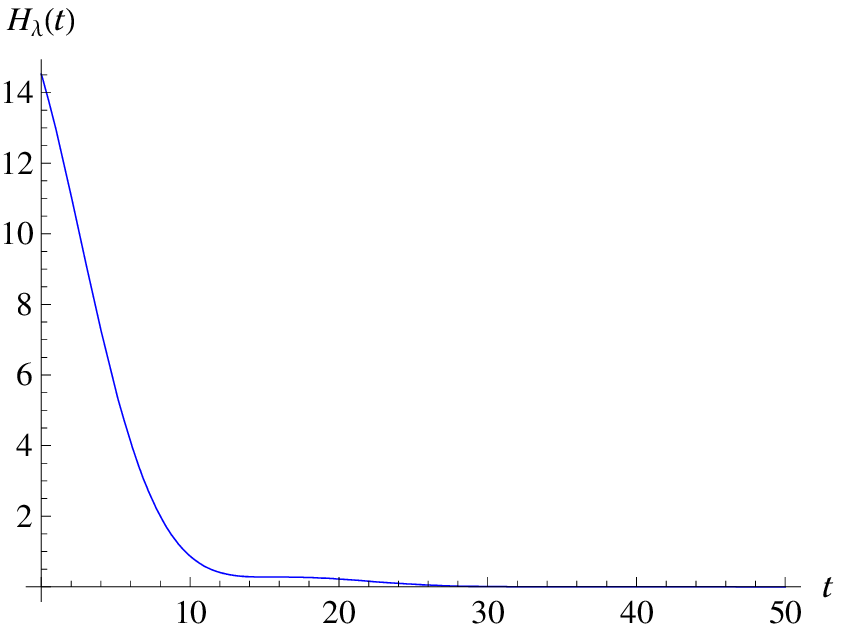}\begin{tabular}{cc}
 & \tabularnewline
\end{tabular}\includegraphics[scale=0.83]{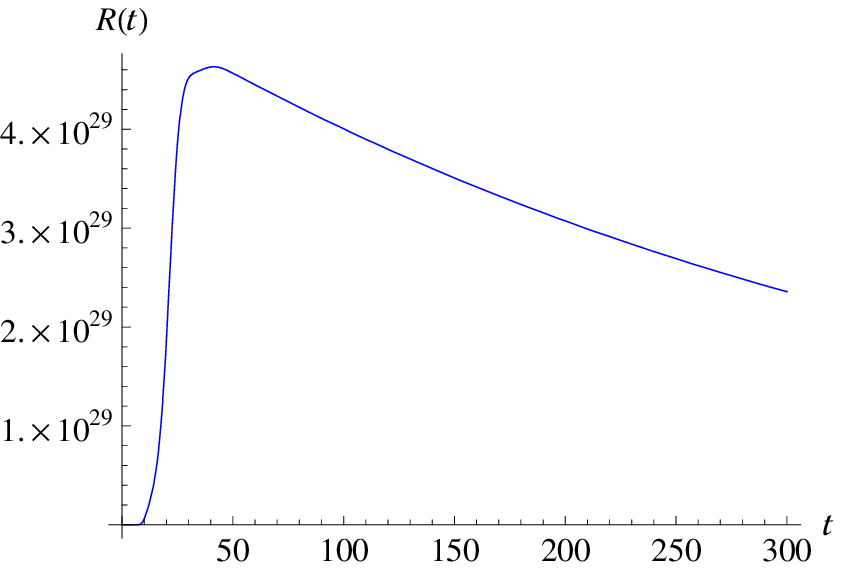}\caption{The end of inflation is before $t\thickapprox\unit[21.3869]{\zeta}$}
\end{figure}

\begin{figure}[H]
\begin{centering}
\includegraphics[scale=0.83]{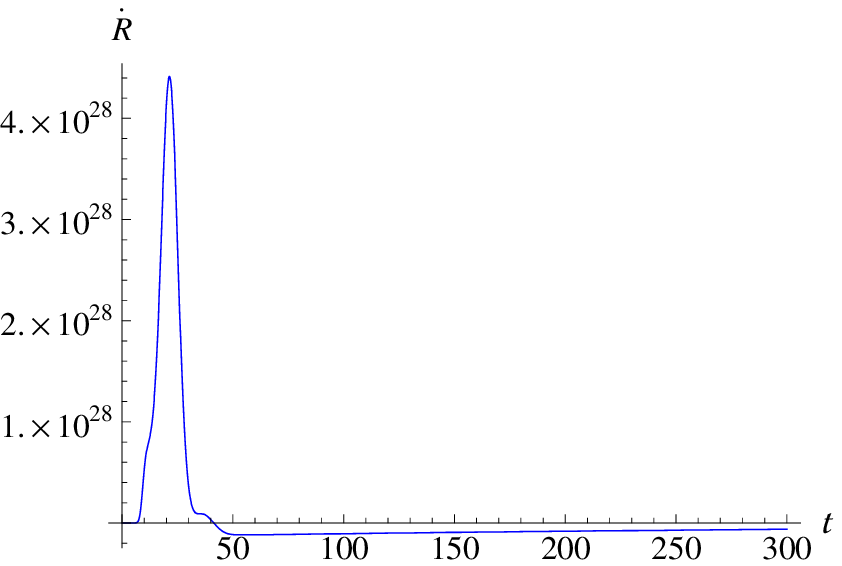}\begin{tabular}{cc}
 & \tabularnewline
\end{tabular}\includegraphics[scale=0.83]{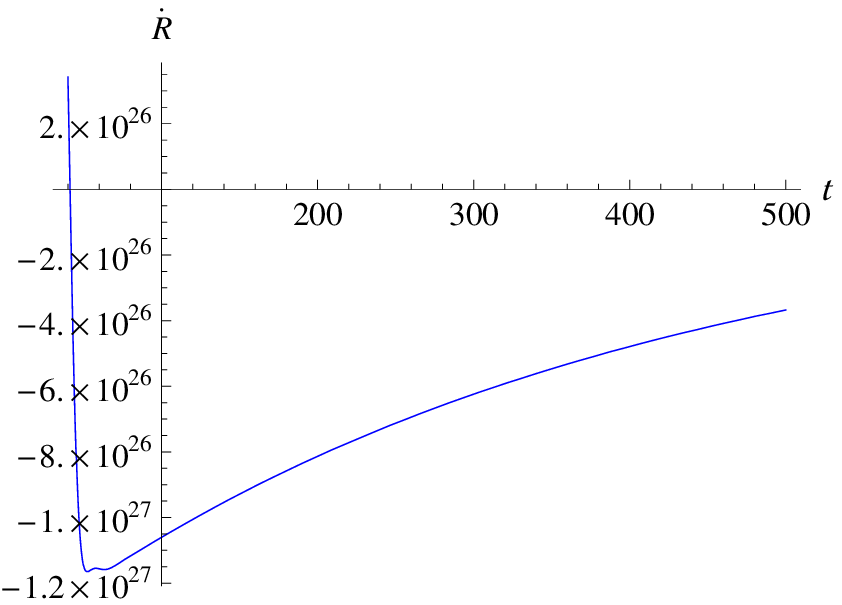}
\par\end{centering}

\caption{We discover $\dot{\mathrm{R}}\left(\unit[41.3895]{\zeta}\right)\thickapprox0$
at $t_{0}\lesssim\unit[41.3896]{\zeta}$ . }
\end{figure}

\begin{figure}[H]
\centering{}\includegraphics[scale=0.83]{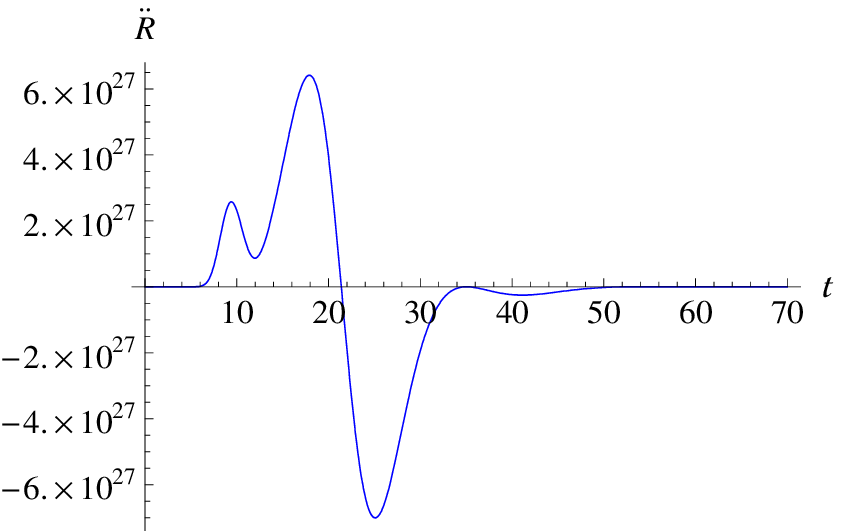}\begin{tabular}{cc}
 & \tabularnewline
\end{tabular}\includegraphics[scale=0.83]{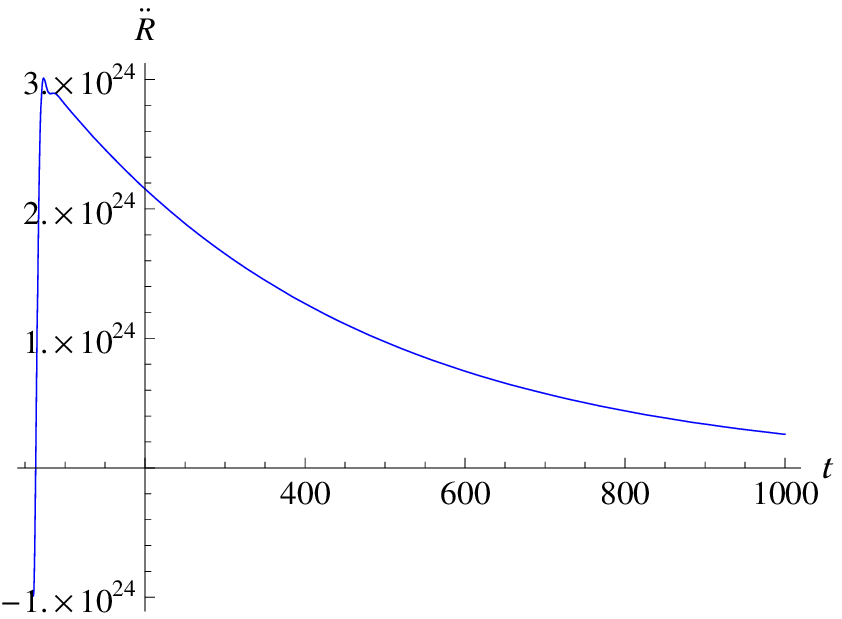}\caption{$\ddot{\mathrm{R}}$ is no longer less than zero after $t\thickapprox\unit[52.4105]{\zeta}$.}
\end{figure}

\begin{figure}[H]
\centering{}\includegraphics[scale=0.83]{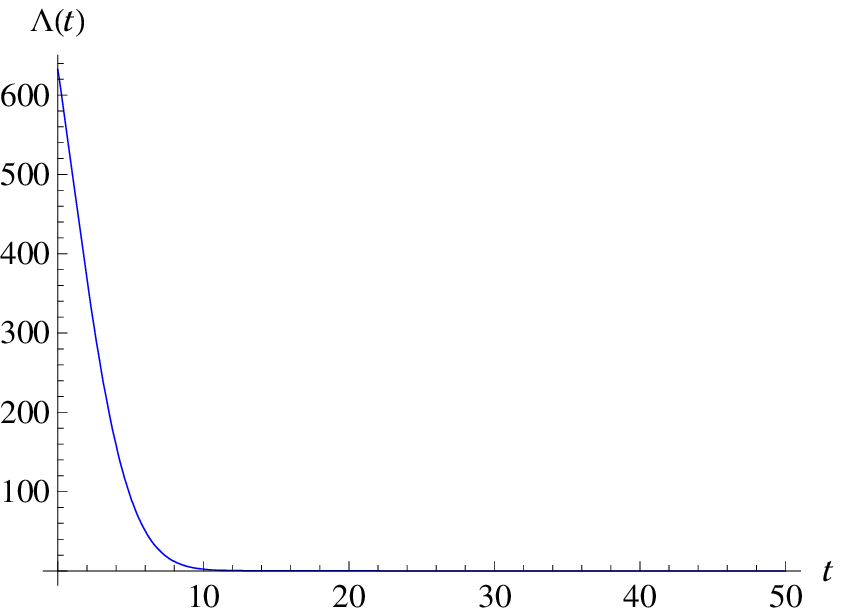}\begin{tabular}{cc}
 & \tabularnewline
\end{tabular}\includegraphics[scale=0.83]{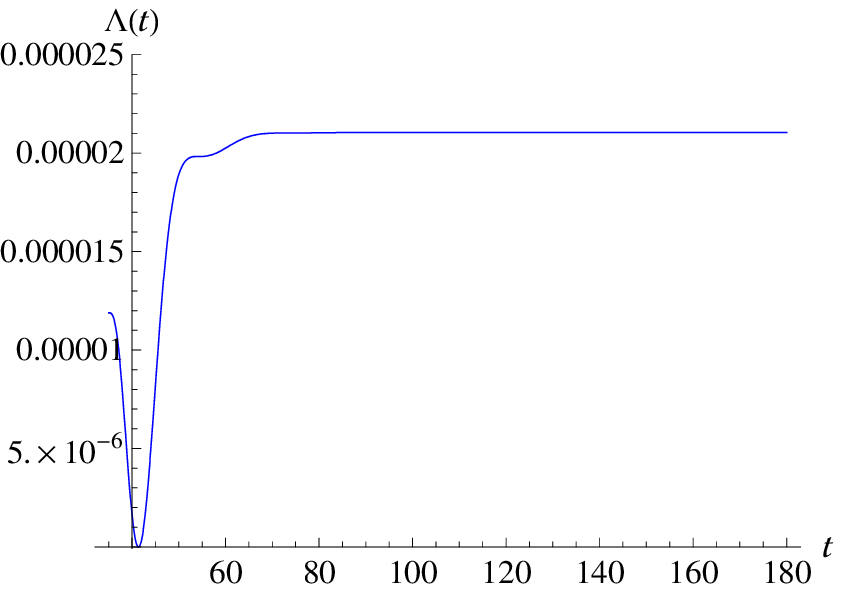}\caption{The cosmological term would become an approximately non-zero positive
constant as $\unit[2.10405\times10^{-5}]{\zeta^{-2}}$ when $t\gtrsim\unit[173.4152]{\zeta}$.}
\end{figure}

\begin{figure}[H]
\centering{}\includegraphics[scale=0.83]{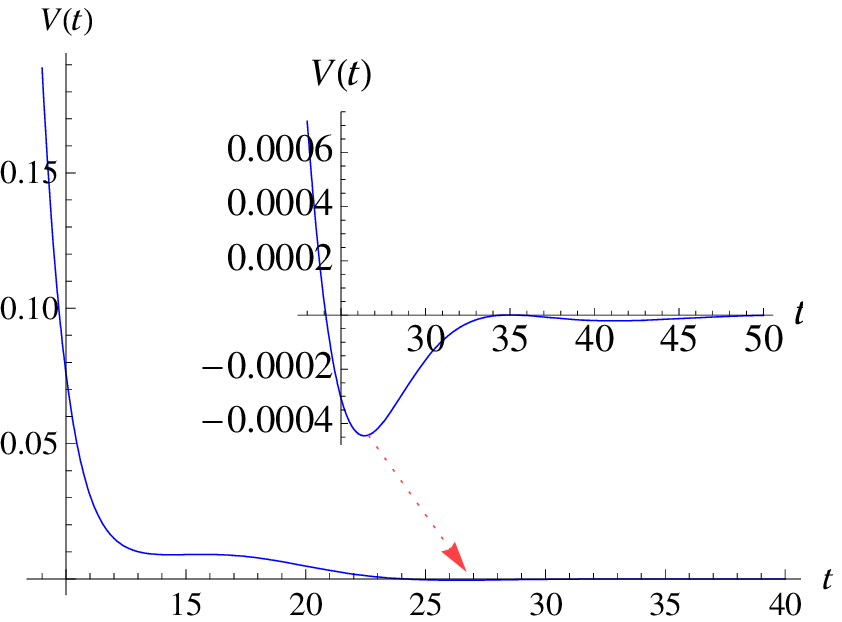}\begin{tabular}{cc}
 & \tabularnewline
\end{tabular}\includegraphics[scale=0.83]{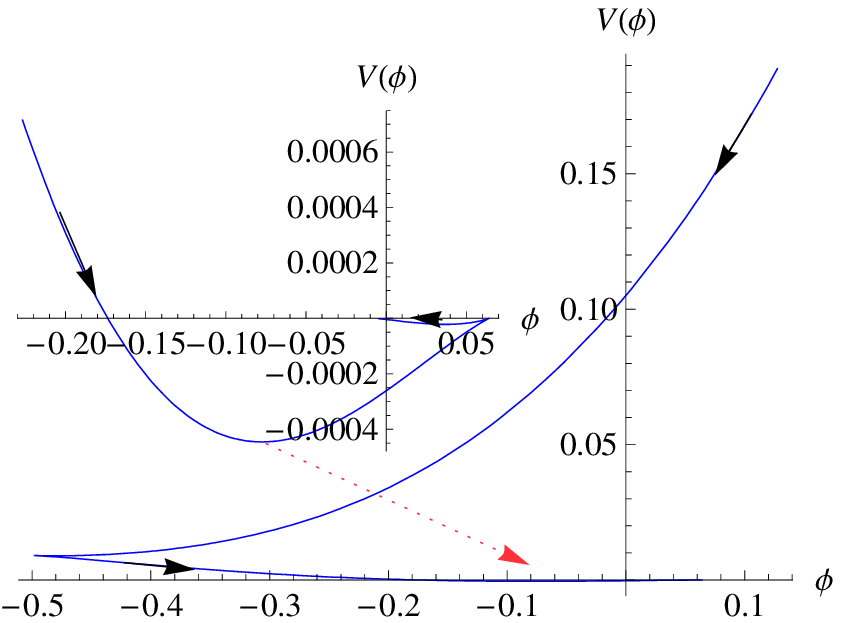}\caption{The time interval of both pictures is $\left[\unit[9]{\zeta},\unit[40]{\zeta}\right]$.
The conditions of the minimum negative potential have been shown by
sub-pictures.}
\end{figure}

\begin{figure}[H]
\centering{}\includegraphics[scale=0.83]{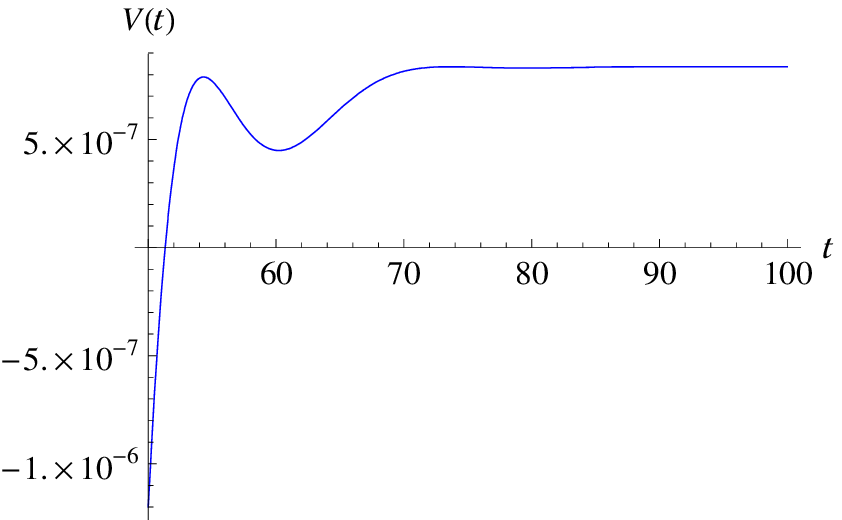}\begin{tabular}{cc}
 & \tabularnewline
\end{tabular}\includegraphics[scale=0.83]{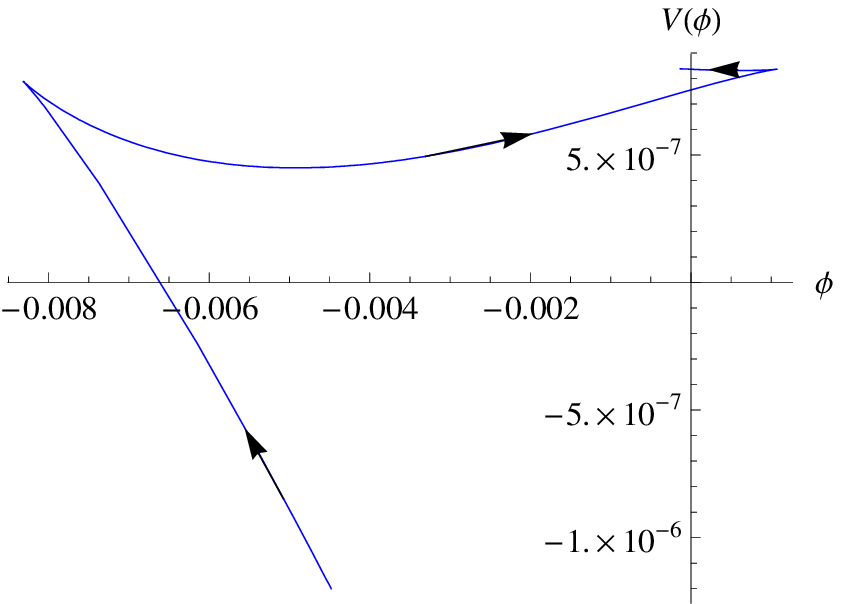}\caption{The time interval of both pictures is $\left[\unit[50]{\zeta},\unit[100]{\zeta}\right]$.
They allow us to see that the potential will rise to positive from
negative, before landing on a positive value after an adequately long
period of time.}
\end{figure}

Certain important data from this example is noteworthy: the period
of inflation is before $\unit[21.387]{\zeta}$; at $t>\unit[41.389]{\zeta}$,
the scale factor begins to shrink, i.e. $\dot{\mathrm{R}}\left(t>\unit[41.389]{\zeta}\right)<0$;
there are three occurrences of negative acceleration during $\unit[21.386]{\zeta}<t\lesssim\unit[63.290]{\zeta}$
before the universe begins to experience positive acceleration, which
gradually slows down the collapse. The value of its potential would
be $\unit[8.37174\times10^{-7}]{\zeta^{-2}}$ when $t\gg\unit[210]{\zeta}$.

\section{discussion and summary}

\subsection*{Analysis of units }

In the previous section, we introduced the units of time {}``$\unit{\zeta}$''
and mass {}``$\unit{M}$'', and fixed them as $\unit{M=\zeta^{-1}}$.
The adoption of these settings contains two benefits: First, we can
choose a proper scale for $\unit{\zeta}$ to fit our inference about
the period of inflation. For example, due to the needs of the discussion
witnessed in \prettyref{eq:12} and \prettyref{eq:27}, we require
the scale factor before the age of the universe reaches $t_{\mathrm{GUT}}\sim\unit[10^{-36}]{s}$
to grow by a factor more than $e^{60}-e^{70}$. Therefore, we can
define $\unit{\zeta}=\unit[10^{5}]{\mathit{t}_{\mathrm{Planck}}}$
$\left(\mathit{t}_{\mathrm{Planck}}\sim\unit[10^{-44}]{s}\right)$
when considering the epoch of inflation at $t_{\mathrm{inf}}\sim\unit[10^{-37}]{s}$.
Of course, the other scale of $\unit{\zeta}$ can be used when inflation
during other eras is explored. As such, corresponding to one\textquoteright{}s
inference, one could, for example, define $\unit{\zeta}=\unit[10^{-3}]{\mathit{t}_{\mathrm{EW}}}$
$\left(\mathit{t}_{\mathrm{EW}}\sim\unit[10^{-11}]{s}\right)$ to
investigate both the epoch of inflation before electroweak phase transition
and the situations that arise from it. Second, a new unit of energy
density can be defined corresponding to \begin{equation}
\unit{\varepsilon_{\mathrm{\zeta}}}=\frac{3c^{2}\zeta^{-2}}{8\pi G}\label{eq:42}\end{equation}
for \prettyref{eq:35}, \prettyref{eq:39}, \prettyref{eq:40} and
\prettyref{eq:41}. Thus, if we adopt $\unit{\zeta}=\unit[10^{5}]{\mathit{t}_{\mathrm{Planck}}}$
and introduce the Planck energy density $\unit{\varepsilon_{\mathrm{Planck}}}=\unit[c^{2}G^{-1}]{\mathit{t}_{Planck}^{-2}}$
$\left(\unit{\varepsilon_{\mathrm{Planck}}}\sim\unitfrac[10^{117}]{GeV}{cm^{3}}\right)$,
we attain $\varepsilon_{\unit{\zeta}}=\unit[\frac{3}{8\pi}\times10^{-10}]{\varepsilon_{\mathrm{Planck}}}$.
Applications of unit-setting for previous tests will be presented
in the appendix.

\subsection*{Data analysis}

According to \prettyref{eq:32}, \prettyref{eq:33} and \prettyref{eq:34},
the evolution of $\int_{t_{i}}^{t}\dot{\phi}^{2}d\tau$ is an important
key for controlling any of the types of universe outlined in Table
I. Therefore, as an universe of type 1 or 2, if it has a non-resting
kinetic scalar field at $t>t_{0}$ that leads \prettyref{eq:36} or
the square root of \prettyref{eq:33} to be negative, this value will
always be negative and the universe will collapse forever, even though
the density of its ordinary matter is extremely thin at the time.
However, if the $\dot{\phi}$ goes to rest quickly enough, causing
\prettyref{eq:34} and \prettyref{eq:36} to both become positive
constants, the universe(s) that we place it in type 4 will be in a
state of accelerating expansion. Additionally, if $\int_{t_{i}}^{t}\dot{\phi}^{2}d\tau$
is sufficiently small for a long enough time to lead \prettyref{eq:36}
to be positive, the universe(s) will be expanding but with uncertain
behavior as in type 5. The reason for this uncertainty is the fact
that we can not have an exact value for $4\pi\dot{\phi}^{2}$ in \prettyref{eq:34}.
To compare, the static type 3 universe(s) as displayed in Table I
would occur with extreme difficulty because a fine-tuned $\dot{\phi}^{2}\left(t\right)$
is needed to make $H\left(t_{i}\right)-4\pi\int_{t_{i}}^{t>t_{r}=t_{0}}\dot{\phi}^{2}d\tau=0$.
This is most unnatural. 

Moreover, we discover that $\ddot{\mathrm{R}}$ of \prettyref{eq:34}
will always be positive after a characteristic time $t_{*}$, even
if/when the universe finally shrinks! This has been verified by our
tests and is expressed in Table I as a collapsed universe of type
1 or 2 (as in figure 11), and an expanded universe of type 4 (as in
figure 6). It looks highly counterintuitive, but a collapsed universe
could only have an epoch of \textquotedblleft{}decelerated collapse\textquotedblright{}
if $t_{*}$ was near to $t_{0}$ and $t_{0}$ big enough to allow
it a very low matter-density. By way of contrast, it seems that the
type 5 universe should not rightly be in Table I because, unlike the
other types of universe, its time period for observation is \emph{before}
the characteristic time $t_{0}$ or $t_{r}$. However, type 5 is absolutely
necessary: without it, our information regarding universes of other
types would be incomplete because we would be unable to investigate
them with either decelerating expansion or both accelerating and decelerating
expansion in rotation.

For explanatory convenience, slightly exaggerated values for the IHPs
are introduced in examples B and C so as to show the accelerating
expansion and collapse of the universe clearly. Following this technique,
we discover that if we wish to have an e-folding number $N\sim60-70$
for the model $\phi\left(t\right)=\phi\left(t_{i}\right)\exp\left(-\frac{t-t_{i}}{\beta}\right)\cos\left[\frac{m}{\sqrt{12\pi}}\left(t-t_{i}\right)\right]$,
$H\left(t_{i}\right)$ must have a value close to $\unit[14.52]{\zeta^{-1}}$
with initial settings in line with those mentioned at the beginning
of Section IV. Actually, such a value seems to be particularly special
because the universe collapses when it is placed at $H\left(t_{i}\right)=\unit[14.518]{\zeta^{-1}}$
and enters a situation of accelerating expansion if it is set as $H\left(t_{i}\right)=\unit[14.523]{\zeta^{-1}}$.
Although these two examples are toy models, a critical value for the
IHP can be determined at about $H\left(t_{i}\right)\gtrsim\unit[14.52064830064115]{\zeta^{-1}}$
for the model as currently proposed. This is according to the conditions
of $N\gtrsim60-70$ and the real cosmological constant $\Lambda\thickapprox\unit[1.934\times10^{-35}]{s^{-2}}$.
Besides, we discover that the circumstances of inflaton mass $m$
and the decay parameter $\beta$ are also essentials for controlling
an universe, regardless of whether it is expanded or collapsed. Indeed,
as outlined above, when $m$ and $\beta$ are fixed, a big enough
value for $H\left(t_{i}\right)$ would make an universe enter accelerating
expansion after inflation. However, a bigger $m$ or $\beta$ with
a fixed $H\left(t_{i}\right)$ would finally lead an universe to collapse.

To enlarge, according to \prettyref{eq:34} and \prettyref{eq:35},
the behavior of $V\left(t\right)$ is analogous to $\nicefrac{\ddot{\mathrm{R}}}{\mathrm{R}}$,
so the potential value will always be positive after a suitably long
time for any universe of types 1, 2 or 4. Of course, the minimum of
$V\left(t\right)$ does not occur at $t>t_{r}$, but at the time when
an universe has a maximum deceleration of $\mathrm{R}\left(t\right)$
in order to stop inflation. This conforms to figure 15 at $t\approx\unit[32]{\zeta}$.
On the other hand, we find that the potential could be a surjective
function of $\phi$ which is according to figures 7 and 8 of example
B and figures 13 and 14 of example C. Actually, the reason for this
is that the potential $V$ should be a function with variables of
$\phi$ and $t$, as seen in \prettyref{eq:16} and \prettyref{eq:31}.
From these figures, we can understand that $V$ would hit a minimum
negative value and then rise to a positive one proportional to $\Lambda$
when $\dot{\phi}$ is finally at rest. This is quite different from
the phase transition model that we are familiar with and it looks
as if it would correspond to the scenario of reheating after inflation.

Additionally, we have found that, from determining the value of the
cosmological constant alone, it is impossible to make conclusions
about the denouement of any universe except a static one. Fortunately,
the $\Lambda-t$ picture does contain an indication that could help
us to differentiate between collapsing and expanding universes though:
as figure 16 displays, there is a rule whereby an universe would be
collapsing if $\Lambda$ increases from lowest point, i.e. zero (as
depicted by the red, solid line), but expanding when $\Lambda$ decreases
smoothly (as depicted by the blue, dotted line).

\begin{figure}[H]
\begin{centering}
\includegraphics[scale=0.83]{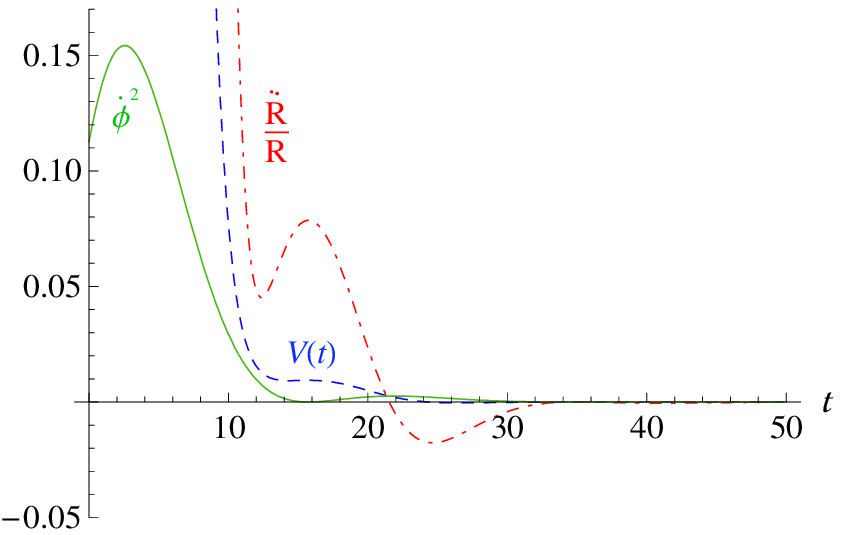}\begin{tabular}{cc}
 & \tabularnewline
\end{tabular}\includegraphics[scale=0.83]{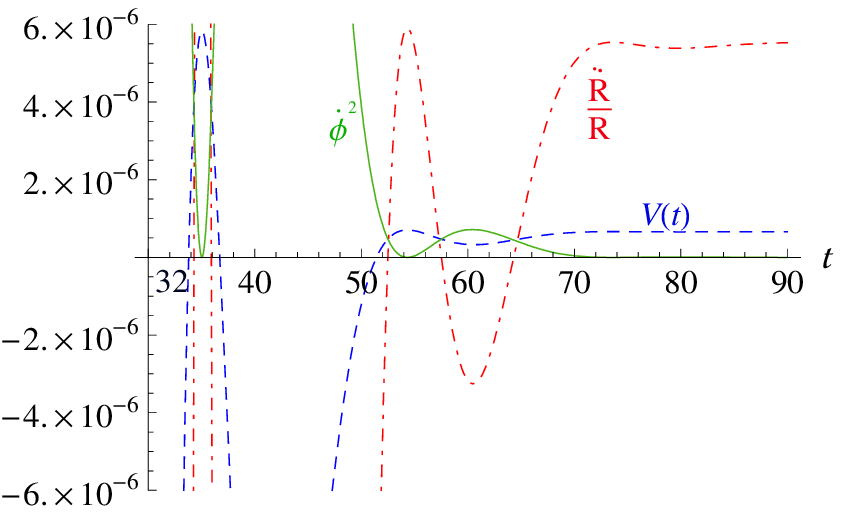}
\par\end{centering}

\centering{}\caption{The above pictures show the behavior of $\nicefrac{\ddot{\mathrm{R}}}{\mathrm{R}}$,
$\dot{\phi}^{2}$ and $V\left(t\right)$ in test B.}
\end{figure}

\begin{figure}[H]
\begin{centering}
\includegraphics[scale=0.83]{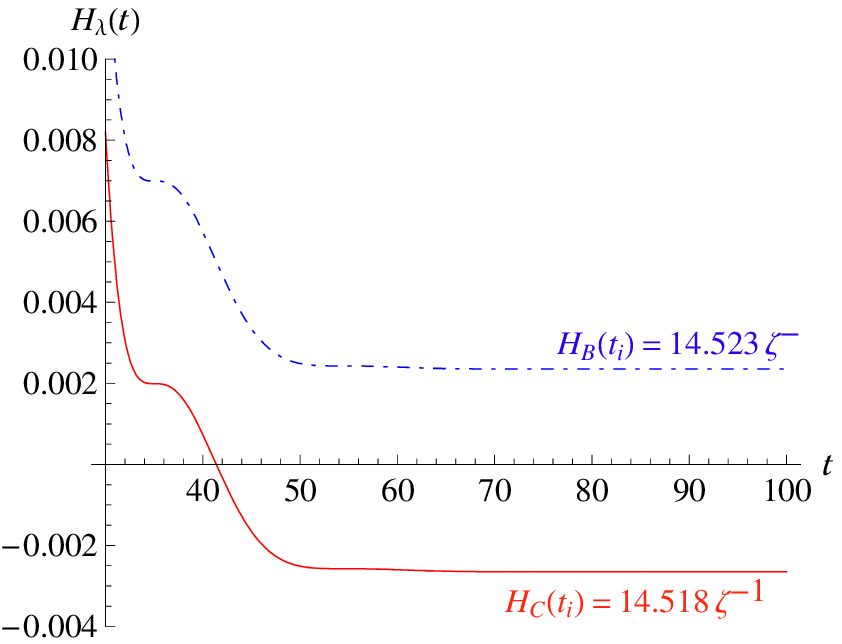}\begin{tabular}{cc}
 & \tabularnewline
\end{tabular}\includegraphics[scale=0.83]{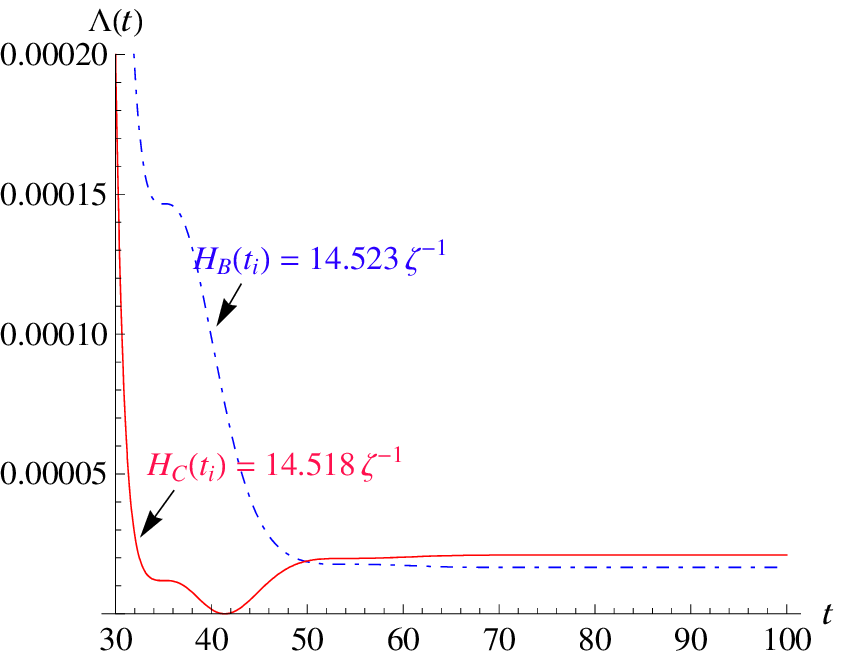}
\par\end{centering}

\caption{These show the behavior of $H_{\lambda}\left(\zeta\right)$ and $\Lambda\left(\zeta\right)$
of type 2 (with a blue, dot-dashed line) and type 4 (with a solid,
red line) universes.}
\end{figure}

\subsection*{Answers to the Aforementioned Problems}

In this article, we have obtained the differential equation \prettyref{eq:31}
according to our assumption of the necessary time-varying scalar field
potential and found that the term of $-3H\dot{\phi}^{2}$ could absorb
energy from the total energy density difference of inflaton. Of course,
we could use the conventional method to calculate \prettyref{eq:28},
\prettyref{eq:29} and \prettyref{eq:30} or \prettyref{eq:31}. This
is done by proposing an ansatz for the potential so as to obtain the
solutions of inflaton and the scale factor of an universe. Through
an analysis of the solutions gained from this process, we are able
to answer corresponding questions about the growth of an universe. 

Irrespective of the above practice, however, we have decided to employ
a new and non-traditional method for solving equations and problems
-- one which gives the ansatz of $\phi\left(t\right)$ that appears
in the solution of the scale factor as \prettyref{eq:32} directly.
Using the scale factor gained in this manner, we can then \textquotedblleft{}rebuild\textquotedblright{}
the potential function as \prettyref{eq:35} according to the ansatz
of $\phi\left(t\right)$. In other words, with the non-traditional
method, potential is replaced by the scalar field for the role of
the ansatz.

The new method does not refute the traditional one, but rather contains
its own advantages: by testing A, B and C, we achieve a feasible method
for analyzing simple solutions of inflaton that are otherwise dependent
on highly complex potential models. Moreover, besides toy-model tests,
the evolution of the inflaton potential energy density (IPED) can
also be easily realized through the careful observation and analysis
of \prettyref{eq:35}. From this, it is clear to see that, apart from
being constrained by $\int_{t_{i}}^{t}\dot{\phi}^{2}d\tau$, the IPED
is also restricted by the initial Hubble parameter $H\left(t_{i}\right)$.
The fact that our proposal enables us to obtain the IHP is a very
important advantage because its corresponding term, as shown in \prettyref{eq:40},
can be viewed as the vacuum energy density of the universe\textquoteright{}s
original situation. Furthermore, we can assert that \prettyref{eq:41}
is Einstein\textquoteright{}s cosmological term, making \prettyref{eq:39}
the effective cosmological constant as needed. 

Thus, we have proposed a scheme that enables a solution to the problem
of the cosmological constant: as long as suitable models of inflaton
are given with a value for the e-folding number as desired, and regardless
of whether these models are based on intuition or observation, we
will be able to obtain the corresponding IHP by inputting the present
observation of the Hubble constant. This then allows solutions for
the evolution of IPED to be uncovered. Finally, we can solve the problems
of the cosmological constant by considering the evolution of the total
energy density of inflaton according to \prettyref{eq:29}, \prettyref{eq:33},
\prettyref{eq:39}, \prettyref{eq:40} and \prettyref{eq:41}.

At this stage, let us quickly review the Friedmann equations in terms
of matter density and pressure with the cosmological constant:

\begin{equation}
\frac{\ddot{\mathrm{R}}}{\mathrm{R}}=-\frac{4\pi G}{3c^{2}}\left(\varepsilon_{\mathrm{matter}}+3p\right)+\frac{\Lambda c^{2}}{3},\label{eq:43}\end{equation}
\begin{equation}
\left(\frac{\mathrm{\dot{R}}}{\mathrm{R}}\right)^{2}=\frac{8\pi G}{3c^{2}}\varepsilon_{\mathrm{matter}}+\frac{\Lambda c^{2}}{3}.\label{eq:44}\end{equation}
Now, \prettyref{eq:43} is adequate for explaining/describing the
re-accelerating expansion of an universe with its consideration of
the cosmological constant\textquoteright{}s existence or/and the negative
pressure of the universe. However, problems ensue when \prettyref{eq:44}
is analyzed: while it yields suitable results for the present moment
in the present universe, it struggles to adequately illustrate a situation
in which a collapse occurs from an expanding universe. In my opinion,
this predicament constitutes a very big loss to the whole theory of
cosmology, especially since we can only explore other kinds of universe
in our imagination and require equations that can help us to do so.
This is why our new, non-conventional method is particularly useful
as it facilitates such theoretical exploration.

Alongside our findings, two additional facts about the evolution of
an universe can be established. First, when an universe appears to
be expanding at time $t$, $H\left(t_{i}\right)$ must be bigger than
the effect of $\int_{t_{i}}^{t}\dot{\phi}^{2}d\tau$; conversely,
when a collapsing universe is considered at $T$, $\int_{t_{i}}^{T}\dot{\phi}^{2}d\tau$
will be bigger than $H\left(t_{i}\right)$, if $\dot{\phi}$ is still
moving at this time. This is a natural conclusion emerging from \prettyref{eq:36}
because $\int_{t_{i}}^{t}\dot{\phi}^{2}d\tau$ grows with time. It
consequently follows that if one can give correct information about
$\dot{\phi}^{2}$ at an arbitrary cosmic time $t$, \prettyref{eq:36}
can be used to describe the evolution of an universe coherently and
not only in an expanding situation, but also in a collapsing one.
For this reason, our proposal is a better alternative than the traditional
method because it gets rid of the predicament that emerges from the
Friedmann equation \prettyref{eq:44}. 

The second fact is as follows: according to \prettyref{eq:35}, we
discover that it could be possible to have a negative value of potential,
such as in the situation of $\dot{\phi}^{2}\left(t_{0}\right)>0$.
We should point out that the negative value appears to disprove our
proposal. Fortunately, however, the catastrophe is averted because
a negative value for the total energy density of inflaton will not
be possible when its dependence on the property of \prettyref{eq:35}
is incorporated. Indeed, the negative potential actually has an advantage
because it combines with $\dot{\phi}^{2}$ in \prettyref{eq:28} to
allow a large enough deceleration for the purpose of ending the period
of inflation. Traditionally, $\dot{\phi}^{2}$ has been problematic
because it requires an extremely specific/fine-tuned value that makes
the universe become the one what we see today. Accordingly, a negative
potential could help $\dot{\phi}^{2}$ to have more possibility during
the end of the inflation. Moreover, the range of running potential
from positive to negative could also help acceleration to smoothly
move between positive and negative as well. This provides an adequate
picture of the universe\textquoteright{}s evolution.

As such, the two facts that we have proposed are thus mechanisms that
can fulfill the necessities of both collapsing and re-accelerating
expanding-type universes. Importantly, we can assert that, with a
proper model $\phi$ and our proposal from \prettyref{eq:33} to \prettyref{eq:41},
the introduction of these two facts will cause the gulf between theories
to disappear.

In reference to our discussion, the appendix provides much important
information about our tests while adopting units with which we are
familiar. It deserves to be mentioned that rows $\dagger$ and $\ddagger$
show the corresponding properties of the current cosmological constant,
which are made by calculating with the approximation of $H\left(t_{i}\right)=\unit[14.52064830064116]{\zeta^{-1}}$
and the same model $\phi$ as with test B.

\section*{appendix }

\begin{table}[H]
\begin{raggedright}
\begin{tabular}{>{\raggedright}m{1.7cm}>{\raggedright}m{3cm}>{\raggedright}m{3cm}>{\raggedright}m{3cm}}
\hline 
 & $t_{e}$ & $t_{0}$ & $t_{*}$\tabularnewline
\hline 
Test A & $9.548\times10^{-38}$ & $1.048\times10^{-37}$ & $1.142\times10^{-37}$\tabularnewline
Test B & $1.161\times10^{-37}$ & none & $3.467\times10^{-37}$\tabularnewline
Test C & $1.153\times10^{-37}$ & $2.231\times10^{-37}$ & $2.826\times10^{-37}$\tabularnewline
\hline
$\dagger$ & $1.157\times10^{-37}$ & none & $1.632\times10^{-37}$\tabularnewline
\hline
\end{tabular}
\par\end{raggedright}

\caption{The unit of time in this table is {}``$\unit{s}$''. Where $t_{e}$
is the time when inflation in the universe has ended; $t_{0}$ is
the time when $\dot{\mathrm{R}}$ becomes negative; and $t_{*}$ is
the time when $\ddot{\mathrm{R}}$ is no longer negative. The unit
of time that we adopted is $\unit{\zeta}=\unit[10^{5}]{\mathit{t}_{\mathrm{Planck}}}$. }
\end{table}
\begin{table}[H]
\begin{raggedright}
\begin{tabular}{>{\raggedright}m{1.8cm}>{\raggedright}m{3.3cm}>{\raggedright}m{3.3cm}>{\raggedright}m{3.3cm}>{\raggedright}m{3.3cm}}
\hline 
 & $\left\langle \varepsilon_{\mathrm{vac}}\right\rangle $ & $\left\langle \varepsilon_{\mathrm{V}}\right\rangle _{initial}$ & $\left\langle \varepsilon_{\mathrm{V}}\right\rangle _{min}$ & $\left\langle \varepsilon_{\mathrm{V}}\right\rangle _{final}=\left\langle \varepsilon_{\Lambda}\right\rangle $\tabularnewline
\hline 
Test A & $1.450\times10^{108}$ & $1.446\times10^{108}$ & $-3.836\times10^{105}$ & none\tabularnewline
Test B & $7.281\times10^{108}$ & $7.265\times10^{108}$ & $-1.133\times10^{104}$ & $1.909\times10^{101}$\tabularnewline
Test C & $7.276\times10^{108}$ & $7.260\times10^{108}$ & $-1.289\times10^{104}$ & $2.421\times10^{101}$\tabularnewline
\hline
$\ddagger$ & $\lesssim7.279\times10^{108}$ & $\lesssim7.263\times10^{108}$ & $\lesssim-1.207\times10^{104}$ & $2.732\times10^{78}$; 

$6.468\times10^{-6}$ $\star$\tabularnewline
\hline
\end{tabular}
\par\end{raggedright}

\centering{}\caption{The unit of energy density is {}``$\unitfrac{GeV}{cm^{3}}$''. We
can calculate the initial energy density by \prettyref{eq:40} and
the minimum potential energy density and energy density of $\Lambda$
by \prettyref{eq:35} and \prettyref{eq:39} respectively. Additionally,
with regard to $\left\langle \varepsilon_{\mathrm{V}}\right\rangle _{initial}=\left\langle \varepsilon_{\mathrm{vac}}\right\rangle -\frac{1}{2}\dot{\phi}^{2}\left(t_{i}\right)$;
$\star$, the upper value is calculated by using $H\left(t_{i}\right)=\unit[14.52064830064116]{\zeta^{-1}}$;
the lower value is obtained through the current cosmological constant
$\Lambda\thickapprox\unit[1.934\times10^{-35}]{s^{-2}}$.}
\end{table}

\begin{acknowledgments}
I would like to thank my best friends Dan and Aleksandar, who gave
me lots support and helped me to modify this article (I particularly
thank Dan as without his help, it wouldn\textquoteright{}t have been
presented in the way I wish it to be). Also, I respectfully appreciate
the suggestions and comments of Prof. Huang, W. Y. and Dr. Gu, J.
A.; the help given by Dr. Liu, T. C.\textquoteright{}s; and, especially,
the discussions with Dr. Cheng, T. C. Importantly, I would further
like to express my gratitude for the support I received from Mr. Bao,
Ms. Lin, Dr. Hsu, Ms. Jane, Ms. Chang, N. Z. and, above all, my parents
and my beautiful wife, Hoki Akiko. Finally, last but not least, I
also wish that my lovely daughter CoCo can receive my thoughts and
gratitude to her. 

To all of the above, I thank you very much. \end{acknowledgments}

\end{document}